%% file: ms.tex
\documentclass[conference]{IEEEtran}
\pdfoutput=1
\IEEEoverridecommandlockouts
\usepackage{booktabs}
\usepackage{url}
\usepackage{balance}
\usepackage{amsmath}
\usepackage{amsfonts}
\usepackage{graphicx}
\usepackage{array}
\usepackage{subcaption}
\usepackage{listings}
\usepackage[most]{tcolorbox}
\usepackage{tikz}
\usetikzlibrary{shapes,positioning,fit}
\usetikzlibrary{arrows}

\def\InnerSep{1.8pt}

\newcounter{theorem2}
\newtheorem{example}[theorem2]{Example}

\usepackage{microtype}
\newcommand{\subparagraph}{}
\usepackage{titlesec}
\titlespacing\section{0pt}{6pt plus 0pt minus 1pt}{0pt plus 0pt minus 1pt}
\titlespacing\subsection{0pt}{5pt plus 0pt minus 1pt}{0pt plus 0pt minus 1pt}
\titlespacing\subsubsection{0pt}{5pt plus 0pt minus 1pt}{0pt plus 0pt minus 1pt}

\begin{document}

\title{A+ Indexes: Tunable and Space-Efficient Adjacency Lists in Graph Database Management Systems}

\author{\IEEEauthorblockN{Amine Mhedhbi, Pranjal Gupta, Shahid Khaliq, Semih Salihoglu}
\IEEEauthorblockA{Cheriton School of Computer Science, University of Waterloo \\
\{amine.mhedhbi, pranjal.gupta, shahid.khaliq, semih.salihoglu\}@uwaterloo.ca}
}

\maketitle

\input{abstract}
\input{introduction}

\input{alists}

\input{implementation}
\input{evaluation}
\input{related-work}
\input{future-work}

\balance

\bibliographystyle{IEEEtran}
\bibliography{references}

\end{document}

%% file: abstract.tex
\begin{abstract}
\label{sec:abstract}
Graph database management systems (GDBMSs) are highly optimized to perform fast traversals, i.e., joins of vertices with their neighbours, by indexing the neighbourhoods of vertices in adjacency lists. However, existing GDBMSs have system-specific and fixed adjacency list structures, which makes each system 
efficient on only a fixed set of workloads. We describe a new tunable indexing subsystem for GDBMSs, we call A+ indexes, with materialized view support. The subsystem consists of two types of indexes: (i) vertex-partitioned indexes that partition 1-hop materialized views into adjacency lists on either the source or destination vertex IDs; and (ii) edge-partitioned indexes that partition 2-hop views into adjacency lists on one of the edge IDs. As in existing GDBMSs, a system by default requires one forward and one backward vertex-partitioned index, which we call the primary A+ index. Users can tune the primary index or secondary indexes by adding nested partitioning and sorting criteria. Our secondary indexes are space-efficient and use a technique we call {\em offset lists}. Our indexing subsystem allows a wider range of applications to benefit from GDBMSs' fast join capabilities. We demonstrate the tunability and space efficiency of A+ indexes through extensive experiments on three workloads.
\end{abstract}

%% file: introduction.tex
\section{Introduction}
\label{sec:introduction}

\noindent The term {\em graph database management system} (GDBMS) in its contemporary usage refers to data management software such as Neo4j~\cite{neo4j}, JanusGraph~\cite{janusgraph}, TigerGraph~\cite{tigergraph}, and GraphflowDB~\cite{mhedhbi:sqs, kankanamge:graphflow} that adopt the property graph data model~\cite{neo4j-property-graph-model}. 
In this model, entities are represented by vertices, relationships are represented by edges, and attributes by arbitrary key-value properties on vertices and edges. GDBMSs have lately gained popularity among a wide range of applications from fraud detection and risk assessment in financial services to recommendations in e-commerce~\cite{sahu:survey}. 
One reason GDBMSs appeal to users is that they are highly optimized to perform very fast joins of vertices with their neighbours. 
This is primarily achieved by using {\em adjacency list indexes}~\cite{bonifati:graphs-book}, which are join indexes that are used by GDBMSs' join operators.

Adjacency list indexes are often implemented using constant-depth data structures, such as the compressed sparse-row (CSR)  structure, that partition the edge records into lists by source or destination vertex IDs. Some systems adopt a second level partitioning in these structures by edge labels. These partitionings provide constant time access to neighbourhoods of vertices and contrasts with tree-based indexes, such as B+ trees, which have logarithmic depth in the size of the data they index. 
Some systems further sort these lists according to some properties, which allows them to use fast intersection-based join algorithms, such as the novel intersection-based worst-case optimal (WCO) join algorithms~\cite{ngo:survey}. 
However, a major shortcoming of existing GDBMSs is that systems make different but fixed choices about the partitioning and sorting criteria of their adjacency list indexes, which makes each system highly efficient on only a fixed set of workloads. This creates physical data dependence, as users have to model their data, e.g., pick their edge labels, according to the fixed partitioning and sorting criteria of their systems. 

We address the following question: {\em How can the fast join capabilities of GDBMSs be expanded to a much wider set of workloads?} We are primarily interested in solutions designed for 
read-optimized GDBMSs. This is informed by a recent survey of users and
applications of GDBMSs that we conducted~\cite{sahu:survey}, that indicated that 
GDBMSs are often used in practice to support read-heavy applications, 
instead of primary transactional stores. 
As our solution, we describe a tunable and space-efficient indexing subsystem for GDBMSs that we call \emph{A+ indexes}.
Our indexing subsystem consists of a {\em primary} index and optional {\em secondary} indexes that users can build. 
This is similar to relational systems that index relations in a primary B+ tree index on 
the primary key columns as well as optional secondary indexes on other columns. 
Primary A+ indexes are the default indexes that store all of the edge records in a database. 
Unlike existing GDMBSs, users can tune the primary A+ index of the system by adding arbitrary 
nested partitioning of lists into sublists and providing a sorting criterion per sublist. We store these lists in
a nested CSR data structure, which provides constant time access
to vertex neighborhoods that can benefit a variety of workloads.

We next observe that partitioning edges into adjacency lists is equivalent to creating multiple materialized views where 
each view is represented by a list or a sublist within a list. Similarly, the union of all adjacency lists can be seen as the 
coarsest view, which we refer to as the {\em global view}. In existing systems and primary A+ indexes, the global view is a 
trivial view that contains all of the edges in the graph. Therefore, one way a GDBMS can support an even wider
range of workloads is by indexing other views inside adjacency lists. 
However storing and indexing views in secondary indexes results in data duplication and 
consumes extra space, which can be prohibitive for some views.

Instead of extending our system with general view functionality, our next contribution carefully identifies 
two sets of global views  
that can be stored in a highly space-efficient manner when partitioned appropriately into lists:
(i) 1-hop views that satisfy arbitrary predicates that are stored in {\em secondary vertex-partitioned A+ indexes}; and (ii) 2-
hop views that are stored in {\em secondary edge-partitioned A+ indexes}, which extend the notion of 
neighborhood from vertices to edges, i.e., each list stores a set of edges that are adjacent to a particular edge.
These two sets of views and their accompanying partitioning methods guarantee that the 
final lists that are stored in secondary A+ indexes are subsets of lists in the primary A+ index.
Based on this property, we implement 
secondary A+ indexes by a technique we call {\em offset lists}, which identify each indexed edge by an 
offset into a list in the primary A+ index. 
Due to the sparsity, i.e., small average degrees, of real-world graphs, each list in the primary A+ index 
often contains a very small number of edges. This makes offset lists highly space-efficient, taking a few bytes per indexed edge instead 
of the ID lists in the primary index that store globally identifiable IDs of edges and neighbor vertices, 
each of  which are often 8 bytes in existing systems. Similar to the primary A+ index, secondary indexes
are implemented in a CSR structure that support nested partitioning, where the lower level is the 
offset lists.
To further improve the space-efficiency of secondary A+ indexes, we identify cases when the secondary A+ indexes can share the partitioning levels of the primary A+ index.    

We implemented A+ indexes inside the GraphflowDB in-memory GDBMS~\cite{kankanamge:graphflow}.
We describe the modifications we made to the optimizer and query processor of the system to  
use our indexes in query plans. 
We present examples of highly efficient plans 
that our system is able to generate using our 
indexing subsystem that do not exist in the plan spaces of existing systems.
We demonstrate the tunability and space efficiency of A+ indexes by showing 
how to tune GraphflowDB to be highly efficient on three different workloads using either primary index reconfigurations or
building secondary indexes with very small memory overhead.
GraphflowDB is a read-optimized system that does
not support transactions but allows non-transactional updates. Although update performance is not our focus,
for completeness of our work, we report the update performance of A+ indexes in the longer version of our paper~\cite{mhedhbi2020a}.

\begin{figure}[t!]
	\centering
	\captionsetup{justification=centering}
	\includegraphics[scale=0.47]{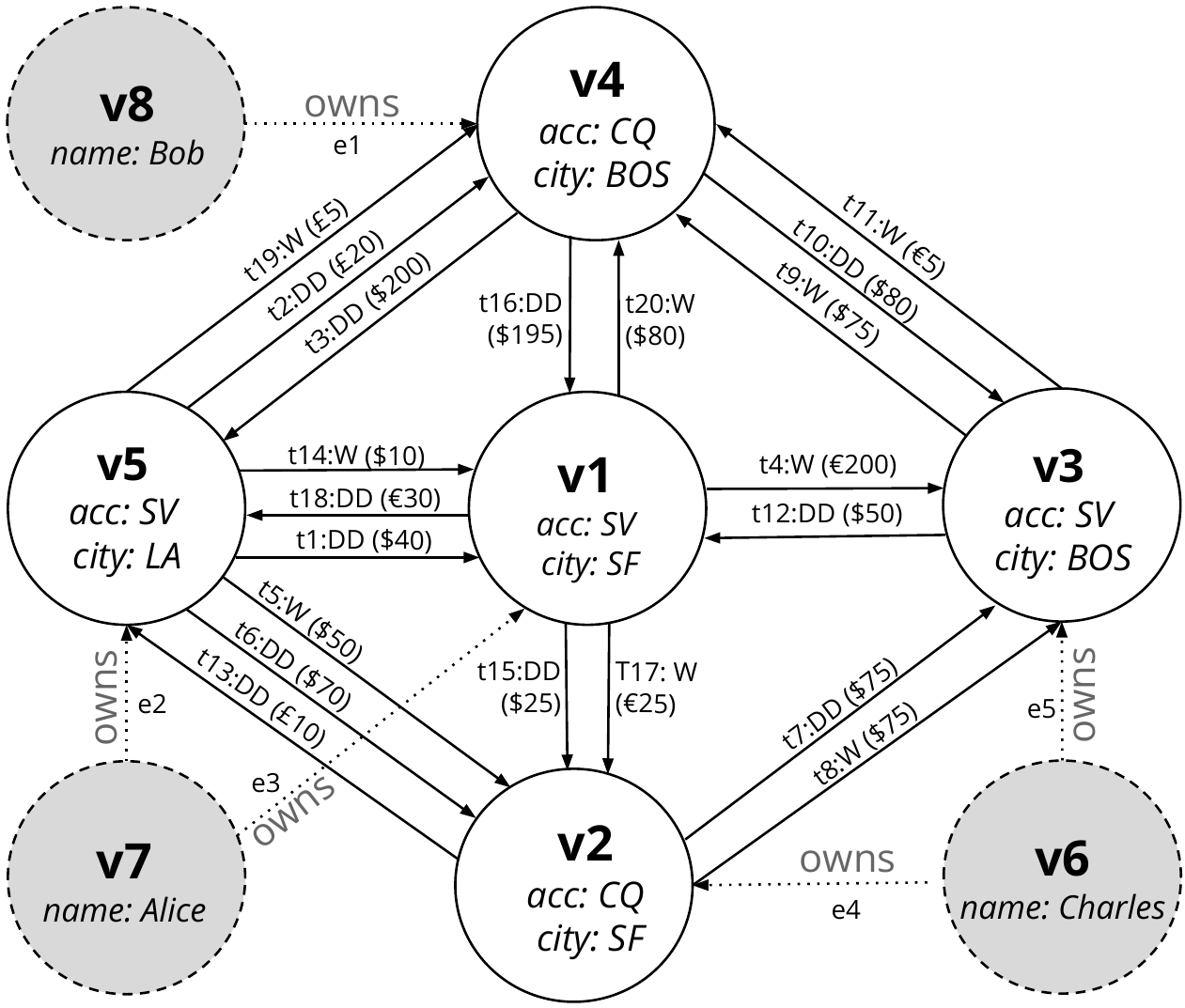}
	\caption{Example financial graph.}
	\label{fig:running-example}
\end{figure}

Figure~\ref{fig:running-example} shows an example financial graph that we use as a running example throughout this paper. The graph contains vertices with \texttt{Customer} and {\ttfamily Account} labels. \texttt{Customer} vertices have \texttt{name} properties and {\ttfamily Account} vertices have \texttt{city} and {\ttfamily accountType(acc)} properties. From customers to accounts are edges with {\ttfamily Owns(O)} labels and between accounts are transfer edges with {\ttfamily Dir-De}\-\texttt{posit(DD)} and {\ttfamily Wire(W)} labels with \texttt{amount(amt)}, \texttt{cu\-rrency}, and {\ttfamily date} properties. We omit dates in the figure and give each transfer edge an ID such that $t_i.date < t_j.date$ if $i < j$.

\input{overview-existing-indexes}

%% file: overview-existing-indexes.tex
\section{Overview of Existing Adjacency List Indexes}
\label{sec:overview-existing-indexes}

Adjacency lists are accessed by GDBMS's join operators e.g., \textsc{Expand} in Neo4j or \textsc{Extend/Intersect} in GraphflowDB, that join vertices with neighbours. GDBMSs employ two broad techniques to provide fast access to adjacency lists while performing these joins:

\noindent {\bf (1) Partitioning:} GDBMSs often partition their edges first by their source or destination vertex IDs, respectively in {\em forward} and {\em backward} indexes; this is the primary partitioning criterion.

\begin{example}
Consider the following 2-hop query, written in open\-Cypher~\cite{openCypher}, that starts from a vertex with name ``Alice''. Below, a$_i$, c$_j$, and r$_k$ are variables for the \texttt{Account} and \texttt{Customer} query vertices and query edges, respectively. 
{\em 
\begin{lstlisting}[numbers=none, mathescape=true, showstringspaces=false]
MATCH $c_1$$-$$[$$r_1$$]$$-$$>$$a_1$$-$$[$$r_2$$]$$-$$>$$a_2$
WHERE $c_1$$.$$n$$a$$m$$e$$\ $=$$`$A$$l$$i$$c$$e$'
\end{lstlisting}
}
\noindent In every GDBMS we know of, this query is evaluated in three steps: (1) scan the vertices and find a vertex with \texttt{name} ``Alice'' and match a$_1$, possibly using an index on the \texttt{name} property. In our example graph, \texttt{v7} would match c$_1$; (2) access \texttt{v7}'s forward adjacency list, often with one lookup, to match c$_1$$\rightarrow$a$_1$ edges; and (3) access the forward lists of matched a$_1$'s to match c$_1$$\rightarrow$a$_1$$\rightarrow$a$_2$ paths. 
\end{example}

\begin{figure}[t!]
	\centering
	\captionsetup{justification=centering}
	\includegraphics[scale=0.52]{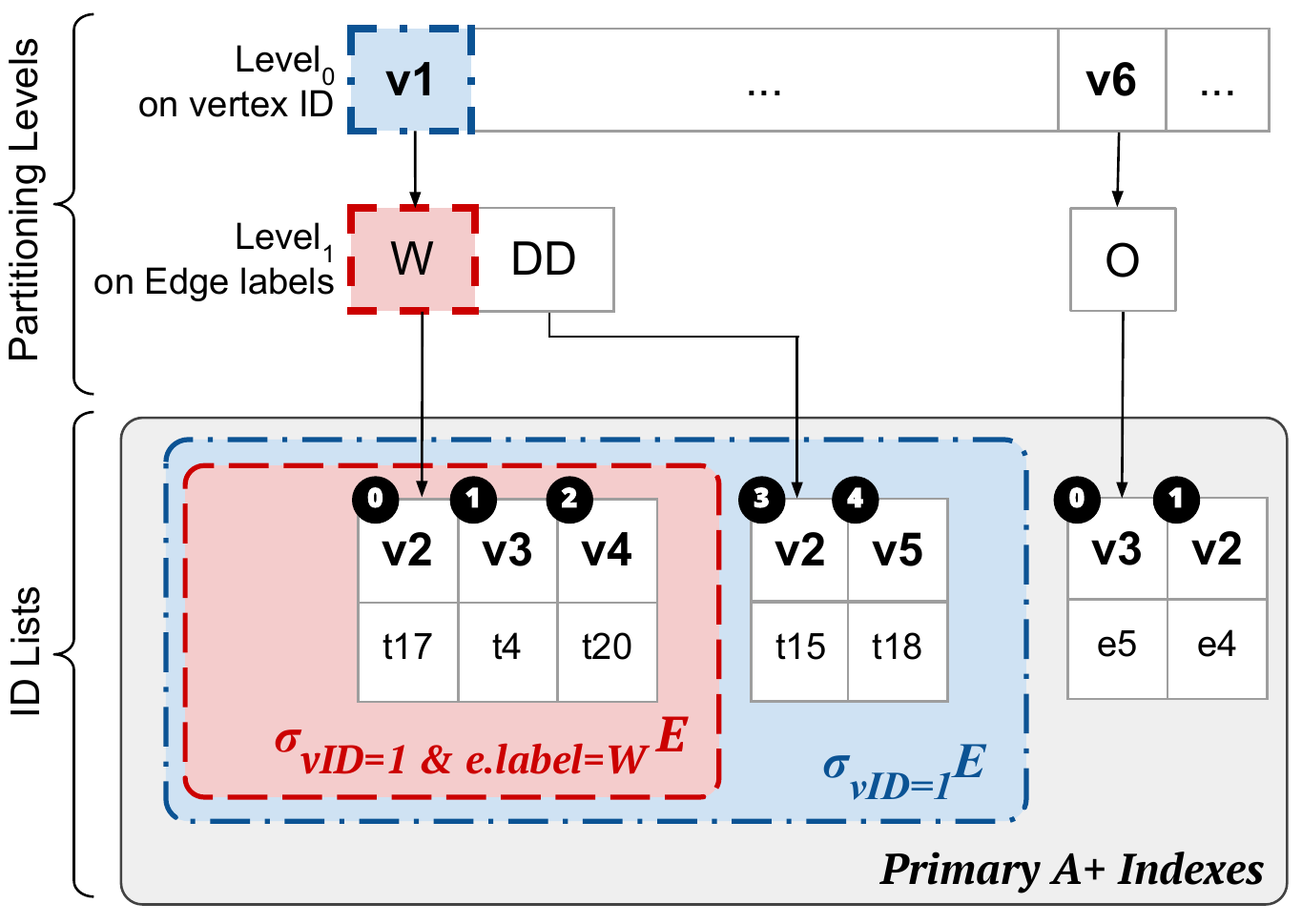}
	\caption{Neo4j Adjacnecy List Indexes.}
	\label{fig:neo4j-example}
\end{figure}

Some GDBMSs employ further partitioning on each adjacency list, e.g., Neo4j~\cite{neo4j} partitions edges on vertices and then by edge labels.  Figure~\ref{fig:neo4j-example} showcases a high-level view of Neo4's paritioning levels and adjacency list index.
Given the ID of a vertex $v$, this allows constant time access to: (i) all edges of $v$; and (ii) all edges of $v$ with a particular label through the lower level lists e.g., all edges of $v$ with label {\texttt Owns}.

\begin{example}
\label{ex:edge-label-partition}
Consider the following query that returns all {\texttt Wire} transfers made from the accounts Alice {\texttt Owns}:
{\em
\begin{lstlisting}[numbers=none, mathescape=true, showstringspaces=false]
MATCH $c_1$$-$$[$$r_1$$:$$O$$]$$-$$>$$a_1$$-$$[$$r_2$$:$$W$$]$$-$$>$$a_2$
WHERE $c_1$$.$$n$$a$$m$$e$$\ $=$$`$A$$l$$i$$c$$e$'
\end{lstlisting}
}
\noindent The ``r$_1$:\texttt{O}'' is syntactic sugar in Cypher for the \texttt{r$_1$.label}$=$ \texttt{Owns} predicate. A system with lists partitioned by vertex IDs and edge labels can evaluate this query as follows. First, find \texttt{v7}, with name ``Alice'', and then access \texttt{v7}'s \texttt{Owns} edges, often with a constant number of lookups and without running any predicates, and match a$_1$'s. Finally access the \texttt{Wire} edges of each a$_1$ to match the a$_2$'s.
\end{example}

\noindent {\bf (2) Sorting:} Some systems further sort their most granular lists  according to an edge property~\cite{janusgraph} or the IDs of the neighbours in the lists~\cite{mhedhbi:sqs, aberger:eh}. Sorting enables systems to access parts of lists in time logarithmic in the size of lists.
Similar to major and minor sorts in traditional indexes, partitioning and sorting keeps the edges in a sorted order, allowing systems to use fast intersection-based join algorithms, such as WCOJs~\cite{ngo:survey} or sort-merge joins.

\begin{example}
\label{ex:alice-cycle}
Consider the following query that finds all 3-edge cyclical wire transfers involving Alice's account $v1$.

\vspace{-4pt}
{\em 
\begin{lstlisting}[numbers=none, mathescape=true, showstringspaces=false]
MATCH $a_1$$-$$[$$r_1$$:$$W$$]$$-$$>$$a_2$$-$$[$$r_2$$:$$W$$]$$-$$>$$a_3$$,$ $a_3$$-$$[$$r_3$$:$$W$$]$$-$$>$$a_1$
WHERE $a_1$$.$ID$=$$v$$1$
\end{lstlisting}
}
\vspace{-4pt}
\noindent In systems that implement worst-case optimal join (WCOJ) algorithms, such as EmptyHeaded~\cite{aberger:eh} or GraphflowDB~\cite{mhedhbi:sqs}, this query is evaluated by scanning each \texttt{v1}$\rightarrow$a$_2$
\texttt{Wire} edge and intersecting the pre-sorted
\texttt{Wire} lists of \texttt{v1} and a$_2$ to match the $a_3$ vertices.
\end{example}

To provide very fast access to each list, lists are often accessed through data structures that have constant depth, such as a CSR instead of logarithmic depths of traditional tree-based indexes. This is achieved by having one level in the index for each partitioning criterion, so levels in the index are not constrained to a fixed size unlike traditional indexes, e.g., k-ary trees. 
Some systems choose alternative implementations. For example Neo4j has a linked list-based
implementation where edges in a list are not stored consecutively but have pointers to each other,
or JanusGraph uses a pure adjacency list design where there is constant time access to all edges of a vertex. 
In our implementation of A+ indexes (explained in Section~\ref{sec:apindexes}), we use CSR as our 
core data structure to store adjacency lists because it is more compact than a pure adjacency list design and achieves better locality than a linked list one. 
Finally, we note that the primary shortcoming of adjacency list indexes in existing systems is that GDBMSs adopt fixed system-specific partitioning and possibly sorting criteria, 
which limits the workloads that can benefit from their fast join capabilities.

%% file: alists.tex
\section{A+ Indexes}
\label{sec:apindexes}
There are three types of indexes in our indexing subsystem: (i) primary A+ indexes; (ii) secondary vertex-partitioned A+ indexes; and (iii) secondary edge-partitioned A+ indexes.
Each index, both in our solution and existing systems, stores a set of adjacency lists, each of which stores a set of edges. We refer to the edges that are stored in the lists as {\em adjacent} edges, and the vertices that adjacent edges point to as {\em neighbour} vertices. 

\subsection{Primary A+ Indexes}
\label{subsec:default-indexes}

The primary A+ indexes are by default the only available indexes. 
Similar to primary B+ tree indexes of relations in relational systems, these indexes are required to contain each edge in the graph, otherwise the system will not be able to answer some queries. Similar to the adjacency lists of existing GDBMSs, there are two primary indexes, one forward and one backward, and we use a nested CSR data structure partitioned first by the source and destination vertex IDs of the edges, respectively.
In our implementation, by default we adopt a second level partitioning by edge labels and sort the most granular lists according to the IDs of the neighbours, which optimizes the system for queries with edge labels and matching cyclic subgraphs using multiway joins computed through intersections of lists. 
However, unlike existing systems, users can reconfigure the secondary partitioning and sorting criteria of primary A+ indexes to tailor the system to variety of workloads, with no or very minimal memory overhead.

\subsubsection{Tunable Nested Partitioning} 
A+ indexes can contain nested secondary partitioning criteria on any categorical property of adjacent edges as well as neighbour vertices, such as edge or neighbour vertex labels, or the \texttt{currency} property on the edges in our running example. In our implementation we allow integers or enums that are mapped to small number of integers as categorical values. 
Edges with null property values form a special partition. 
 
 \begin{example}
	\label{ex:USD}
	Consider querying all wire transfers made in USD currency from Alice's account and the destination accounts of these transfers:
{\em
\begin{lstlisting}[numbers=none, mathescape=true, showstringspaces=false]
 MATCH $c_1$$-$$[$$r_1$$:$$O$$]$$-$$>$$a_1$$-$$[$$r_2$$:$$W$$]$$-$$>$$a_2$ 
 WHERE $c_1$$.$$n$$a$$m$$e$$\ $=$$`$A$$l$$i$$c$$e$'$,$ $r_2$$.$$c$$u$$r$$r$$e$$n$$c$$y$$=$$U$$S$$D$
\end{lstlisting}
}
\noindent Here the query plans of existing systems that partition by edge labels will read all \texttt{Wire} edges from Alice's account and, for each edge, read its \texttt{currency} property and run a predicate to verify whether or not it is in USD. 
\end{example}

Instead, if queries with equality predicates on the \texttt{currency} property are important and frequent for an application, users can reconfigure their primary A+ indexes to provide a secondary partitioning based on \texttt{currency}. 
\begin{lstlisting}[numbers=none, mathescape=true, showstringspaces=false]
RECONFIGURE PRIMARY INDEXES
PARTITON BY $e_{adj}$$.$$l$$a$$b$$e$$l$$,$ $e_{adj}$$.$$c$$u$$r$$r$$e$$n$$c$$y$ 
SORT BY $v_{nbr}$$.$$c$$i$$t$$y$
\end{lstlisting}
\noindent In index creation and modification commands, we use reserved keywords $e_{adj}$ and $v_{nbr}$ to refer to adjacent edges and neighbours, respectively. The above command (ignore the sorting for now) will reconfigure the primary adjacency indexes to have two levels of partitioning after partitioning by vertex IDs:  first by the edge labels and then by the \texttt{currency} property of these edges. For the query in Example~\ref{ex:USD}, the system's join 
operator can now first directly access the lowest level partitioned lists of Alice's list, first by \texttt{Wire} and then by \texttt{USD}, without running any predicates.

\begin{figure}[t!]
	\begin{subfigure}{0.47\textwidth}
		\centering
		\captionsetup{justification=centering}
		\includegraphics[scale=0.54]{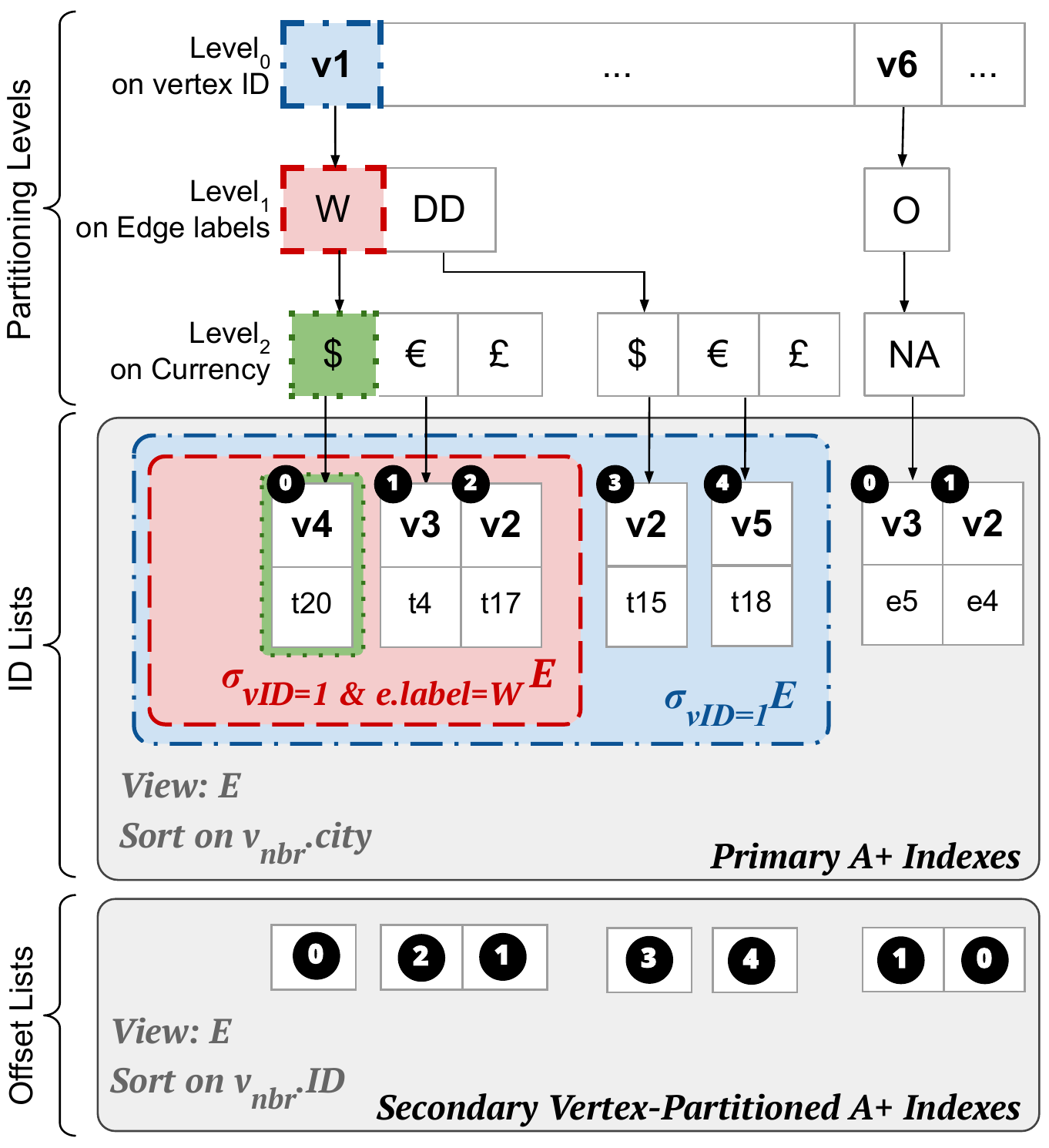}
		\vspace{-2pt}
		\caption{Example primary adjacency lists and secondary vertex-partitioned adjacency lists.}
		\label{fig:global-view}
	\end{subfigure}\vspace{2pt}
	\begin{subfigure}{0.47\textwidth}
		\centering
		\captionsetup{justification=centering}
		\includegraphics[scale=0.54]{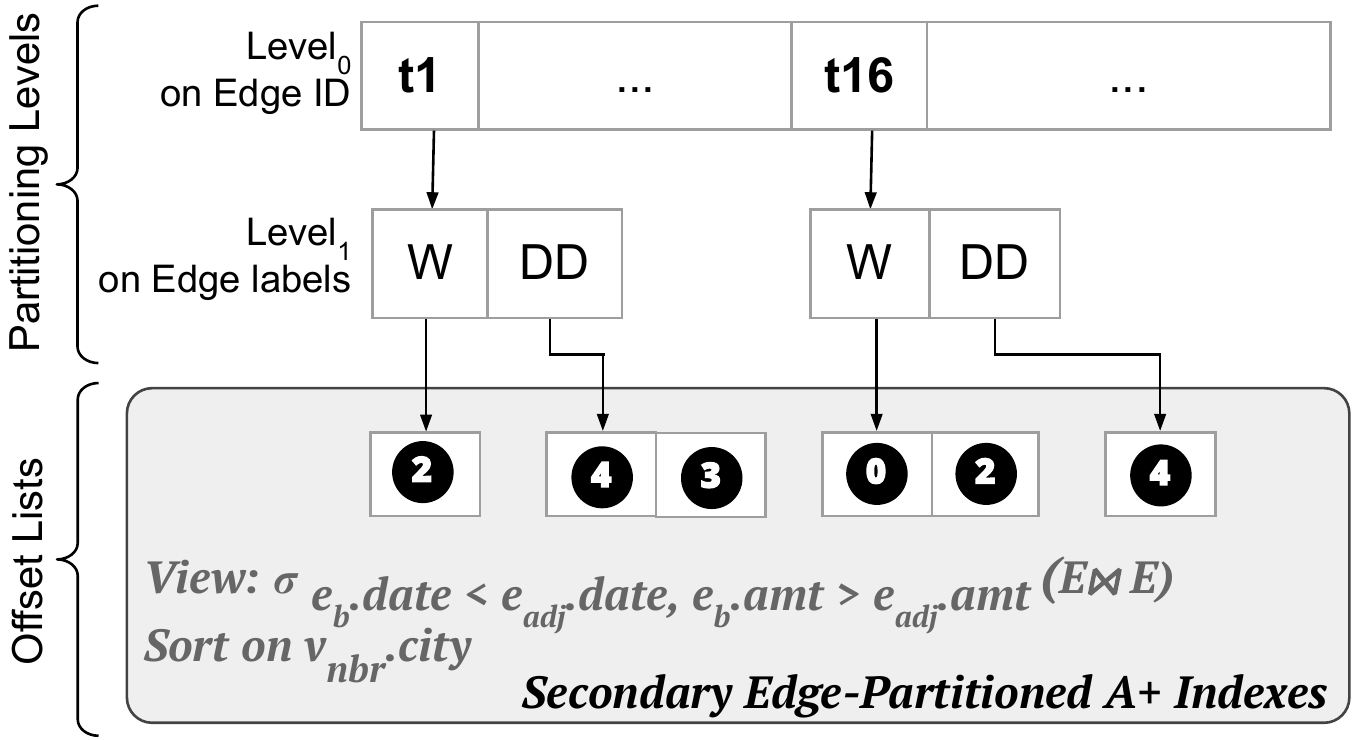}
		\vspace{-2pt}
		\caption{Example secondary edge-partitioned A+ Index.}
		\label{fig:global-view2}
	\end{subfigure}
	\caption{Example A+ indexes on our running example.}
	\label{fig:aplus-indexes}
	\vspace{5.5pt}
\end{figure}

Figure~\ref{fig:global-view} shows the final physical design this generates as an example on our running example.  
We store primary indexes in nested CSR structures. 
Each provided nested partitioning adds a new partitioning level to the CSR, storing offsets to a particular slice of the next layer. After the partitioning levels, at the lowest level of the index are {\em ID lists}, which store the IDs of the edges and neighbour vertices. 
The ID lists are a consecutive array in memory that contains a set of nested sublists. For example, consider the second 
level partitions of the primary index in Figure~\ref{fig:global-view}. Let $L_W$, $L_{DD}$, and $L$ be the list of \texttt{Wire}, \texttt{Dir-Deposit}, and all edges of a vertex $v$, respectively. Then within $L$, which is the list between indices 0-4, are sub-lists $L_W$ (0-2) and $L_{DD}$ (3-4), i.e., $L=L_W \cup L_{DD}$.

\subsubsection{Tunable ID List Sorting} 
The most granular sublists can be sorted according to one or more arbitrary properties of the adjacent edges or neighbour vertices, e.g., the \texttt{date} property of \texttt{Transfer} edges and the \texttt{city} property of the \texttt{Account} vertices of our running example. Similar to partitioning, edges with null values on the sorting property are ordered last. Secondary partitioning and sorting criteria together store the neighbourhoods of vertices in a particular sort order, allowing a system to generate WCOJ intersection-based plans for a wider set of queries.

\begin{example}
\label{ex:ordering-index}
Consider the following query that searches for a three-branched money transfer tree, consisting of wire and direct deposit transfers, emanating from an account with \texttt{vID v5} and ending in three sink accounts in the same city. 
{\em 
\begin{lstlisting}[numbers=none, mathescape=true, showstringspaces=false]	
MATCH $a_1$$-$$[$$:$$W$$]$$-$$>$$a_2$$-$$[$$:$$W$$]$$-$$>$$a_3$, $a_1$$-$$[$$:$$W$$]$$-$$>$$a_4$
      $a_1$$-$$[$$:$$D$$D$$]$$-$$>$$a_5$$-$$[$$:$$D$$D$$]$$-$$>$$a_6$
WHERE $a_1$$.$ID$=$$v5$$,$ $a_3$$.$$c$$i$$t$$y$$=$$a_4$$.$$c$$i$$t$$y$$=$$a_6$$.$$c$$i$$t$$y$ 
\end{lstlisting}
}
If \texttt{Wire} and \texttt{Dir-Deposit} lists are partitioned or sorted by \texttt{city}, as in the above reconfiguration command, after matching
 a$_1$$\rightarrow$a$_2$ and  a$_1$$\rightarrow$a$_5$, a plan can directly intersect two \texttt{Wire} 
 lists of a$_1$ and $a_2$ and one  \texttt{Dir-Deposit} list of $a_5$ in a single operation to find the flows that end up in accounts in the same city. Such plans are not possible with the adjacency list indexes of existing systems.
\end{example}

Observe that the ability to reconfigure the system's primary A+ indexes provides more physical data independence. Users do not have to model their datasets according to the system's default physical design and changes in the workloads can be addressed simply with index reconfigurations. 

\subsection{Secondary A+ Indexes}
Many indexes in DBMSs can be thought of as data structures that give fast access to {\em views}.  In our context, each sublist in the primary indexes is effectively a {\em view} over edges. 
For example, the red dashed list in Figure~\ref{fig:global-view} is the $\sigma_{\text{srcID=v1 \&e.label=Wire}}$\texttt{Edge} view while the green dotted box encloses a more selective view corresponding to $\sigma_{\text{srcID=1 \& e.label=wire \& curr=USD}}$\texttt{Edge}. Each nested sublist in the lowest-level ID lists is a view with one additional equality predicate.
One can also think of the entire index as indexing a {\em global view}, which for primary indexes is simply the \texttt{Edge} table. Therefore the views that can be obtained through the system's primary A+ index are constrained to views over the edges that contain an equality predicate on the source or destination ID (due to vertex ID partitioning) and one equality predicate for each secondary partitioning criterion.  

To provide access to even wider set of views, a system should support more general materialized views and index these in adjacency list indexes. However, supporting additional views and materializing them inside additional adjacency list indexes requires data duplication and storage. 
We next identify two classes of global views and ways to partition these views
that are conducive to a space-efficient implementation: (i) 1-hop views that are stored in secondary vertex-partitioned A+ 
indexes; and (ii) 2-hop views that are stored in secondary edge-partitioned A+ indexes. 
These views and partitioning techniques generate lists that are subsets of the lists in the primary index, which 
allows us to store them in space-efficient {\em offset lists} that exploit 
the small average-degree of real-world graphs and use a few bytes per indexed edge. 
In Sections~\ref{subsec:secondary-vb-indexes} and~\ref{subsec:secondary-eb-indexes} we first describe our logical
views and how these views are partitioned into lists. Similar to the primary A+ index, these lists are stored in CSR-based structures. Section~\ref{sec:offset-lists} describes our offset list-based storage and how we can further increase the space efficiency of secondary A+ indexes by avoiding the partitioning levels of the CSR structure when possible.

\subsubsection{Secondary Vertex-Partitioned A+ Indexes: 1-hop Views}
\label{subsec:secondary-vb-indexes}

Secondary vertex-partitioned indexes store 1-hop views, i.e., 1-hop queries, that contain arbitrary selection predicates on the edges and/or source or destination vertices of edges. These views cannot contain other operators, such as group by's, aggregations, or projections, so their outputs are a subset of the original edges.
Secondary vertex-partitioned A+ indexes store these 1-hop views first by partitioning on vertex IDs (source or destination) and then by the further partitioning and sorting options provided by the primary A+ indexes. In order to use secondary vertex-partitioned A+ indexes, users need to first define the 1-hop view, and then define the partitioning structure and sorting criterion of the index.
\vspace{0.2em}
\begin{example}
\label{ex:secondary-vb}
Consider a fraud detection application that searc\-hes money flow patterns with high amount of transfers, say over 10000 USDs. We can create a secondary vertex-partitioned index to store those edges in lists, partitioned first by vertices and then possibly by other properties and in a sorted manner as before.
{\em
\begin{lstlisting}[numbers=none, mathescape=true, showstringspaces=false]
CREATE 1-HOP VIEW LargeUSDTrnx
MATCH $v_{s}$$-$$[$$e_{adj}$$]$$-$$>$$v_{d}$
WHERE $e_{adj}$$.$$c$$u$$r$$r$$e$$n$$c$$y$$=$$U$$S$$D$$,$ $e_{adj}$$.$$a$$m$$t$$>$$10000$
INDEX AS FW-BW
PARTITION BY $e_{adj}$$.$$l$$a$$b$$e$$l$  SORT BY $v_{nbr}$$.$$I$$D$
\end{lstlisting}
}
\end{example}
\noindent Above, $v_s$ and $v_d$ are keywords to refer to the source and destination vertices, whose properties can be accessed in the WHERE clause. FW and BW are keywords to build the index in the forward or backward direction, a partitioning option given to users. FW-BW indicates indexing in both directions.
The inner-most (i.e., most nested) sublists of the resulting index materializes a view of the form $\sigma_{\text{srcID=* \& elabel=* \& curr=USD}}$ $_{\text{\& amount $>$ 10000}}$\texttt{Edge}. If such views or views that correspond to other levels of the index appear as part of queries, the system can directly access these views in constant time and avoid evaluating the predicates in these views. 

\subsubsection{Secondary Edge-Partitioned A+ Indexes: 2-hop Views}
\label{subsec:secondary-eb-indexes}
Secondary edge-partitioned indexes store 2-hop views, i.e., results of 2-hop queries. As before, these views cannot contain other operators, such as group by's, aggregations, or projections, so their outputs are a subset of 2-paths. The view has to specify a predicate and that predicate has to access properties of both edges in 2-paths (as we momentarily explain, otherwise the index is redundant). Secondary edge-partitioned indexes store these 2-hop views first by partitioning {\em on edge IDs} and then, as before, by the same partitioning and sorting options provided by the primary A+ indexes.
Vertex-partitioned indexes in A+ indexes and existing systems provide fast access to the adjacency of a vertex given 
the ID of that vertex.  Instead, 
our edge-partitioned indexes provide fast access to the {\em adjacency of an edge} given the ID of that edge.
This can benefit applications in which the searched patterns concern relations between two adjacent, i.e., consecutive, edges. We give an example:

\begin{example}
\label{ex:money-flow}
	Consider the following query, which is the core of an important class of queries in financial fraud detection.
{\em
\begin{lstlisting}[numbers=none, mathescape=true, showstringspaces=false]
MATCH $a_1$$-$$[$$r_1$$:$$$$]$$-$$>$$a_2$$-$$[$$r_2$$:$$$$]$$-$$>$$a_3$$-$$[$$r_3$$:$$$$]$$-$$>$$a_4$
WHERE $r_1$$.$eID$=$$t13$$,$
$r_1$$.$$d$$a$$t$$e$$<$$r_2$$.$$d$$a$$t$$e$  $\&$ $r_2$$.$$a$$m$$t$$<$$r_1$$.$$a$$m$$t$$<$$r_2$$.$$a$$m$$t$$+$$\alpha$  $\&$ 
$r_2$$.$$d$$a$$t$$e$$<$$r_3$$.$$d$$a$$t$$e$ $\&$ $r_3$$.$$a$$m$$t$$<$$r_2$$.$$a$$m$$t$$<$$r_3$$.$$a$$m$$t$$+$$\alpha$
\end{lstlisting}
}
\noindent The query searches a three-step money flow path from a transfer edge with \texttt{eID t13} where each additional transfer (\texttt{Wire} or \texttt{Dir-Deposit}) happens at a later date and for a smaller amount of at most $\alpha$, simulating some money flowing through the network with intermediate hops taking cuts.
\end{example}
The predicates of this query compare properties of an edge on a path with the previous edge on the same path. Consider a system that matches r$_1$ to \texttt{t13}, which is from vertex \texttt{v2} to \texttt{v5}. Existing systems have to read transfer edges from \texttt{v5} and filter those that have a later \texttt{date} value than \texttt{t13} and also have the appropriate \texttt{amount} value.
Instead, when the next query edge to match r$_2$ has predicates depending on the query edge r$_1$, these queries can be  evaluated much faster if adjacency lists are partitioned by edge IDs:  a system can directly access the {\em destination-forward adjacency list of} \texttt{t13} in constant time, i.e., edges whose \texttt{srcID} are \texttt{v5}, that satisfy the predicate on the \texttt{amount} and \texttt{date} properties that depend on \texttt{t13}, and perform the extension. 
Our edge-partitioned indexes allow the system to generate plans that perform this much faster processing.
Note that in an alternative design we can partition the same set of 
2-hop paths by vertices instead of edges. However, this would store the same number of edges but
would be less efficient during query processing.
To see this, suppose a system first matches $r_1$ to the edge $(v_2)$$-$$[$\texttt{t13}$]$$-$$>$$(v_5)$ 
and consider extending this edge. The system
can either extend this edge by one more edge to $r_2$, which would require looking up the 2-hop edges
of $v_2$ and find those that have \texttt{t13} as the first edge. This is slower than directly looking up
the same edges using \texttt{t13} in an edge-partitioned list.
Alternatively the system can extend \texttt{t13} by two more edges to 
$[$$r_2$$]$$-$$>$$a_3$$-$$[$$r_3$$]$$-$$>$$a_4$ by accessing the 2-hop edges
of $v_5$ but would need to run additional predicates to check if the edge matching $r_2$ satisfy the necessary
predicates with \texttt{t13}, so effectively processing all 2-paths of $v_5$ and running additional predicates,
which are also avoided in an edge-partitioned list.

There are three possible 2-paths, $\rightarrow\rightarrow$, $\rightarrow\leftarrow$, and $\leftarrow\leftarrow$. Partitioning these paths by different edges gives four unique possible ways in which an edge's adjacency can be defined:

\begin{enumerate}
\item Destination-FW: $v_s$$-$$[e_b]$$\rightarrow$$v_d$$-$$[e_{adj}]$$\rightarrow$$v_{nbr}$
\item Destination-BW: $v_s$$-$$[e_b]$$\rightarrow$$v_d$$\leftarrow$$[e_{adj}]$$-$$v_{nbr}$
\item Source-FW: $v_{nbr}$$-$$[e_{adj}]$$\rightarrow$$v_s$$-$$[e_b]$$\rightarrow$$v_d$
\item Source-BW: $v_{nbr}$$\leftarrow$$[e_{adj}]$$-$$v_s$$-$$[e_b]$$\rightarrow$$v_d$
\end{enumerate}
$e_b$, for ``bound'', is the edge that the adjacency lists will be partitioned by, and $v_{s}$ and $v_{d}$ refer to the source and destination vertices of $e_b$, respectively. For example, the Destination-FW adjacency lists of edge $e$($s$,$d$) stores the forward edges of d. To facilitate the fast processing described above  
for the money flow queries in Example~\ref{ex:money-flow}, we can create the following index:
 
\begin{lstlisting}[numbers=none, mathescape=true, showstringspaces=false]
CREATE 2-HOP VIEW MoneyFlow
MATCH $v_s$$-$$[$$e_b$$]$$\rightarrow$$v_d$$-$$[$$e_{adj}$$]$$\rightarrow$$v_{nbr}$
WHERE $e_b$$.$$d$$a$$t$$e$$<$$e_{adj}$$.$$d$$a$$t$$e$$,$ $e_{adj}$$.$$a$$m$$t$$<$$e_b$$.$$a$$m$$t$
INDEX AS PARTITION BY $e_{adj}$$.$$l$$a$$b$$e$$l$ SORT BY $v_{nbr}$$.$$c$$i$$t$$y$
\end{lstlisting}

The location of the variable $e_b$ in the query implicitly defines the type of partitioning, which in this example is Destination-FW.
This query creates an index that, for each edge $t_i$, stores the forward edges from $t_i$'s destination vertex which have a later \texttt{date} and a smaller \texttt{amount} than $t_i$, partitioned by the labels of their adjacent edges and sorted by the \texttt{city} property of the neighbouring vertices, i.e., the vertex that is not shared with $t_i$. Figure \ref{fig:global-view2} shows the lists this index stores on our running example. 
The inner-most lists in the index correspond to the view: 
$\sigma_{\text{e$_b$.ID=* \& e$_{adj}$.label=* \& e$_b$.date $<$ e$_{adj}$.date \& e$_b$.amt $>$ e$_{adj}$.amt}}(\rho_{e_b}(E) \bowtie$\\$\rho_{e_{adj}}(E))$.
\texttt{E} abbreviates \texttt{Edge} and the omitted join predicate is e$_b$.dstID$=$e$_{adj}$.srcID. Readers can verify that, in presence of this index, a GDBMS can evaluate the money flow query from Example 4 (ignoring the predicate with \texttt{$\alpha$}) by scanning only one edge. It only scans \texttt{t13}'s list which contains a single edge \texttt{t19}. In contrast, even if all \texttt{Transfer} edges are accessible using a vertex-partitioned A+ index, a system would access 9 edges after scanning \texttt{t13}.

Observe that unlike vertex-partitioned A+ indexes, an edge $e$ in the graph can appear in multiple adjacency lists in an edge-partitioned index. For example, in Figure \ref{fig:global-view2}, edge \texttt{t17} (having offset \texttt{2}) appears both in the adjacency list for \texttt{t1} as well as \texttt{t16}. As a consequence, when defining edge-partitioned indexes, users have to specify a predicate that accesses
properties of both edges in the 2-hop query.
This is because if all the predicates are only applied to a single query edge, say $v_s$$-$$[e_b]$$\rightarrow$$v_d$, then we would redundantly generate duplicate adjacency lists. Instead, defining a secondary vertex-partitioned A+ index would give the same access path to the same lists without this redundancy. 

Consider the following example: 

\begin{lstlisting}[numbers=none, mathescape=true, showstringspaces=false]
CREATE 2-HOP VIEW Redundant
MATCH $v_s$$-$$[$$e_b$$]$$\rightarrow$$v_d$$-$$[$$e_{adj}$$]$$\rightarrow$$v_{nbr}$
WHERE $e_{adj}$$.$$a$$m$$t$$<$$1$$0$$0$$0$$0$
\end{lstlisting}

\noindent In absence of an INDEX AS command, views are only partitioned by edge IDs. Consider the account \texttt{v2} in our running example graph in Figure~\ref{fig:running-example}. For each of the four incoming edges of \texttt{v2}, namely \texttt{t5}, \texttt{t6}, \texttt{t15}, and \texttt{t17}, this 
index would contain the same adjacency list that consists of all outgoing edges of \texttt{v2}: $\{$\texttt{t7},\texttt{t8}, \texttt{t13}$\}$,  because the predicate is only on a single edge. Instead, a user can define a vertex-partitioned A+ index with the same predicate and \texttt{v2}'s list would provide an access path to the same edges $\{$\texttt{t7},\texttt{t8},\texttt{t13}$\}$. 

We further note that although we will describe a space-efficient physical implementation of these indexes momentarily, 
the total number of edges in edge-partitioned indexes can be as large as the sum of the squares of degrees unless 
a selective predicate is used, which can be prohibitive for an in-memory system. In our evaluations, we will assume 
a setting where a selective enough predicate is used. 
For 2-hop views that do not have selective predicates, a system should resort to partial materialization of 
these views to reduce the memory consumption under user-specified levels. Partial materialized views is a technique from 
relational systems that has been introduced in reference~\cite{zhou2005partially}, where parts of the view is materialized and others are 
evaluated during runtime. We have left the integration of this technique to future work.

%% file: implementation.tex
\subsubsection{Offset List-based Storage of Secondary A+ Indexes}
\label{sec:offset-lists}
The predominant memory cost of primary indexes is the 
storage of the IDs of the adjacent edges and neighbour vertices.
Because the IDs in these lists globally identify vertices and edges, their sizes need to be logarithmic in the number of edges and vertices in the graph, and are often stored as 4 to 8 byte integers in systems. For example, in our implementation, edge IDs take 8 and neighbour IDs take 4 bytes.
%

In contrast, the lists in both secondary vertex- and edge-partitioned indexes have an important property, which can be exploited to reduce their memory overheads: they are subsets of some ID list in the primary indexes.
Specifically,
a list $L_v$ that is bound to $v_i$ in a secondary vertex-partitioned index is a subset of one of $v_i$'s ID lists. A list $L_e$ that is bound to $e=(v_s, v_d)$ in a secondary edge-partitioned index is a subset of either $v_s$'s or $v_d$'s primary list, depending on the direction of the index, e.g., $v_d$'s list for a Destination-FW list. 
Recall that in our CSR-based implementation, the ID lists of each vertex are contiguous. Therefore, instead of storing an (edge ID, neighbour ID) pair for each edge, we can store a single offset to an appropriate ID list. We call these lists \emph{offset lists}. 
The average size of the ID lists is proportional to the average degree in the graph, which is often very small, in the order of tens or hundreds, in many real world graph data sets. This 
important property of real world graphs has two advantages: 
\begin{enumerate}
\item Offsets only need to be list-level identifiable and can take a small number of bytes which is much smaller than a globally identifiable (edge ID, neighbour ID) pair.
\item Reading the original (edge ID, neighbour ID) pairs through offset lists require an indirection and lead to reading not-necessarily consecutive locations in memory. However, because the ID list sizes are small, we still get very good CPU cache locality. 
\end{enumerate}
An alternative implementation design here is to use a bitmap instead of offset lists. A bitmap
can identify whether each edge in the lists of the primary A+ index is a secondary A+ index. 
This design has the shortcoming that it cannot support the cases when the sorting criterion of secondary A+ indexes is
different than the primary index. However when the sorting criteria are the same, 
this is also a reasonable design point. This has the advantage that when the predicates 
in the lists are not very selective, bitmaps can be even more compact than offset lists, as they require
a single bit for each edge. However reading the edges would now require additional bitmask operations. 
In particular, irrespective of the actual number of edges stored in a secondary index, the system would 
need to perform as many  bitmask operations as the number of edges in the lists of the primary index.  
Therefore as predicates in secondary indexes get more selective, bitmaps would progressively lose
their storage advantage over offset lists and at the same time progressively perform worse in terms of access time.

We implement each secondary index in one of two possible ways, depending on whether the index contains any predicates and whether its partitioning structure matches the secondary structure of the primary A+ indexes.
\begin{itemize}
\item \emph{With no predicates and same partitioning structure}: In this case, the only difference
between the primary and the secondary index is the final sorting of the edges. Specifically, both indexes
have identical partitioning levels, with identical CSR offsets, and the same set of edges in each
inner-most ID/offset sublists, but they sort these sublists in a different order. Therefore we can use the 
partitioning levels of the primary index also to access the lists of the secondary index and save space.
Figure~\ref{fig:global-view} gives an example. The bottom offset lists are for a secondary vertex-partitioned index that has the same partitioning structure as the primary index but sorts on neighbors' IDs instead of neighbors' \texttt{city} properties. Recall that since edge-partitioned indexes need to contain predicates between adjacent edges, this storage can only be used for vertex-partitioned indexes. 
\item \emph{With predicates or different partitioning structure}: In this case, the inner-most sublists of the indexes may contain different sets of edges, so the CSR offsets in the partitioning levels of the primary index cannot be reused 
and we store new partitioning levels as shown in Figure~\ref{fig:global-view2}. 
\end{itemize}
\noindent We give the details of the memory page structures that store ID and offset lists in Section~\ref{sec:other-implementation}. 
\input{implementation-details}

%% file: implementation-details.tex
\section{Implementation Details}
\label{sec:other-implementation}


We implemented our indexing subsystem in GraphflowDB~\cite{mhedhbi:sqs, kankanamge:graphflow} and describe our changes to the system to enable the use of A+ indexes for fast join processing.

\subsection{Query Processor, Optimizer and Index Store}
\label{subsec:processor-optimizer}
A+ indexes are used in evaluating subgraph pattern component of queries, which is where the queries' joins are described. 
We give an overview of the join operators that use A+ indexes and the optimizer of the system. Reference~\cite{mhedhbi:sqs} describes the details of the \textsc{Extend/Intersect} operator and the DP join optimizer of the system in absence of A+ indexes.

\noindent 
{\bf \textsc{Join Operators}:} \textsc{Extend/Intersect (E/I)} is the primary join operator of the system.
Given a query $Q(V_Q, E_Q)$ and an input graph $G(V, E)$,
let a {\em partial k-match} of $Q$ be a set of vertices
of V assigned to the projection of $Q$ onto a set of $k$ query vertices. We denote
a sub-query with $k$ query vertices as $Q_k$. 
\textsc{E/I} is configured to intersect $z$$\ge$$1$ adjacency lists that are 
sorted on neighbour IDs. The operator takes as input
(k-1)-matches of $Q$, performs a $z$-way intersection, and extends them by a single query vertex to k-matches. 
For each (k-1)-match $t$, the operator intersects $z$ adjacency lists of the matched vertices in $t$ and extends $t$ with each vertex in the result of this intersection to produce $k$-matches. If $z$ is one, no intersection is performed, and the operator simply extends $t$ to each vertex in the adjacency list.
The system uses \textsc{E/I} to generate plans that contain 
WCOJ multi-way intersections.
	

To generate plans that use A+ indexes, we first extended \textsc{E/I} to take adjacency lists that can be partitioned by edges as well as vertices. 
We also added a variant of \textsc{E/I} that we call  {\em \textsc{Multi-Extend}}, that performs intersections of adjacency lists that are sorted by properties other than neighbour IDs and extends partial matches to more than one query vertex. 

\noindent {\bf Dynamic Programming (DP) Optimizer and \textsc{Index Store}:} GraphflowDB has a DP-based join optimizer
that enumerates queries one query vertex at a time~\cite{mhedhbi:sqs}. We extended the system's 
optimizer to use A+ indexes as follows.
For each $k$$=$$1,...,m$$=$$|V_Q|$, in order, the optimizer finds the lowest-cost plan for each sub-query $Q_k$ in two ways: (i) by considering extending every possible sub-query $Q_{k-1}$'s (lowest-cost) plan by an \textsc{E/I} operator; and (ii) if $Q$ has an equality predicate involving $z$$\ge$$2$ query edges, by considering extending smaller sub-queries $Q_{k-z}$ by a \textsc{Multi-Extend} operator. 
At each step, the optimizer considers the edge and vertex labels and other predicates together,
since secondary A+ indexes may be indexing views that contain predicates other than edge label equality.
When considering possible $Q_{k-z}$ to $Q_k$ extensions, the optimizer queries the
\textsc{Index Store} to find both vertex- and edge-partitioned indexes, $I_1, ..., I_t$, that 
can be used. 
\textsc{Index Store} maintains the metadata of each A+ index in the system such as their type, partitioning structure, and sorting criterion, as well as additional predicates for secondary indexes.
An index $I_{\ell}$ can potentially be used in the extension if
the edges in the lists in a level $j$ of $I_{\ell}$ satisfy two conditions: (i) extend 
partial matches of $Q_{k-z}$ to $Q_k$, i.e., can be bound to
a vertex or edge in $Q_{k-z}$ and match a subset of the query edges in $Q_z$; and (ii)
the predicates  $p_{\ell, j}$ satisfied in these lists  subsume the predicate $p_Q$ (if any) that is part of this extension.
We search for two types of predicate subsumption. First is conjunctive 
predicate subsumption. If both $p_{\ell, j}$  and $p_Q$ are conjunctive predicates,
we  check if each component of $p_{\ell, j}$ matches a component of $p_Q$. 
Second is range subsumption. If $p_Q$ and $p_{\ell, j}$ or one of their components are range predicates comparing a property against a constant, 
e.g., $e_{adj}$$.$$a$$m$$t$$>$$15000$ and $e_{adj}$$.$$a$$m$$t$$>$$10000$, respectively, we check
if the range in  $p_{\ell, j}$ is less selective than $p_Q$.

Then for each possible index combination retrieved, the optimizer enumerates a plan for $Q_k$ with: (i)  
an \textsc{E/I} or \textsc{Multi-Extend} operator; and (ii) possibly a \textsc{Filter} operator if there are any predicates
that are not fully satisfied during the extension (e.g., if  $p_{\ell, j}$ and $p_Q$ are conjunctive but  $p_{\ell, j}$ does not satisfy all components of $p_Q$). If  the $Q_{k-z}$ to $Q_k$ extension requires using 
multiple indices, so requires performing an intersection, then the optimizer also checks that the sorting criterion on the indices that are returned are the same. Otherwise, it discards this combination.
The systems' cost metric is \emph{intersection cost} (i-cost), which is the total
estimated sizes of the adjacency lists that will be accessed by the \textsc{E/I} and \textsc{Multi-Extend} operators in a plan. 

We note that our optimizer extension to use A+ indexes is 
similar to the classic {\em System R-style} approach to
enumerate plans that use views composed of select-project-join queries directly in 
a DP-based join optimizer~\cite{halevy:survey, chaudhuri:mvs}.  
This approach
also performs a bottom up DP enumeration of join orders of a SQL query $Q$ and
for a sub-query $Q'$ of $Q$, considers evaluating $Q'$ by joining a smaller $Q''$ with a view $V$.
The primary difference is that GraphflowDB's join optimizer
enumerates plans for progressively larger queries that contain, in relational terms, one more column 
instead of one more table (see reference~\cite{mhedhbi:sqs} for details).
Other GDBMSs that use bottom up join optimizers can be extended in a similar way if 
they implement A+ indexes. For example, Neo4j also uses a mix of DP and greedy bottom up enumerator~\cite{neo4j}
called {\em iterative DP}, which is based on reference~\cite{kossman:idp}. However,
extending the optimizers of GDBMSs that use other techniques might require other approaches, e.g., 
RedisGraph, which converts Cypher queries into GraphBLAS linear algebra expression~\cite{graphblas} 
and optimizes this expression.
We also note that we implemented a limited form of predicate subsumption checking.
The literature on query optimization using views 
contains more  general techniques for logical implication of predicates between queries and views~\cite{chaudhuri:mvs, goldstein:mv, ullman:dbkbs}, e.g., detecting that $A > B$ and $B > C$ imply $A > C$.
These techniques can
enhance our implementation and we have not integrated
such techniques within the scope of our paper.

\subsection{Details of Physical Storage}
Primary and secondary vertex-partitioned A+ indexes are implemented using a CSR for groups of 64 vertices and allocates one data page for each group. Vertex IDs are assigned consecutively starting from 0, so given the ID of $v$, with a division and mod operation we can access the second partitioning level of the index storing CSR offsets of $v$. 
The CSR offsets in the final partitioning level point to either ID lists in the case of the primary A+ indexes or offset lists in the case of secondary A+ indexes. The neighbour vertex and edge ID lists are stored as 4 byte integer and 8 byte long arrays, respectively. In contrast, the offset lists in both cases are stored as byte arrays by default. Offsets are fixed-length and use the maximum number of bytes needed for any offset across the lists of the 64 vertices, i.e. it is the logarithm of the length of the longest of the 64 lists rounded to the next byte.


\subsection{Index Maintenance}
\label{sec:index-maintenance}
Each vertex-partitioned data page, storing ID lists or offset lists, is accompanied with an  update buffer. Each edge addition  $e$$=$$(u, v)$ is first applied to the update buffers for $u$'s and $v$'s pages in the primary indexes. Then we go over each secondary vertex-partitioned A+ index $I_V$ in the \textsc{Index Store}. If $I_V$ indexes a view that contains a predicate $p$, we first apply $p$ to see if $e$ passes the predicate. If so, or if $I_V$ does not contain a predicate, we update the necessary update buffers for the offset list pages of $u$ and/or $v$. The update buffers are merged into the actual data pages when the buffer is full. Edge deletions are handled by adding a ``tombstone'' for the location of the deletion until a merge is triggered.

Maintenance of an edge-partitioned A+ index $I_E$ is more involved. For an edge insertion $e$$=$$(u,v)$, we perform two separate operations. First, we check to see if $e$ should be inserted into the adjacency list of any adjacent edge $e_b$ by running the predicate $p$ of $I_E$ on $e$ and $e_b$. 
For example, if $I_E$ is defined as Destination-FW, we loop through all the backward adjacent edges of $u$ using the system's primary index. This is equivalent to running two \emph{delta-queries} as described in references~\cite{kankanamge:graphflow, ammar:bigjoin} for a continuous 2-hop query. Second, we create a new list for $e$
and loop through another set of adjacency lists (in our example $v$'s forward adjacency list in \texttt{D}) and insert edges into $e$'s list. 

\subsection{Index Selection}
\label{sec:index-selection}
Our work focuses
on the design and implementation of a tunable indexing subsystem so that users can tailor a GDBMS to be highly efficient on a wide range of workloads. However, an important aspect of any DBMS is to help users pick indexes from a space of indexes that can benefit their workloads. Given a workload $W$, the space of A+ indexes that can benefit $W$ can be enumerated by enumerating each 1-hop and 2-hop sub-query $Q'$ of each query $Q$ in $W$ and identifying the equality predicates on categorical properties of these sub-queries, which are candidates for partitioning levels, and non-equality predicates on other properties, which are candidates for sorting criterion (any predicate is also a candidate predicate of a global view). 
Given a workload $W$ and possibly a 
space budget $B$, one approach from prior literature to automatically select a subset of these candidate indices
that are within the space budget $B$ 
is to perform a ``what if'' index  simulation to
see the effects of this candidate indices on the estimated costs of plans. 
For example, this general approach is used in  Microsoft SQL Server's 
AutoAdmin~\cite{what-if} tool. 
We do not focus on the problem of recommending a subset of these indexes to users. There are several prior work on index and materialized view recommendation~\cite{what-if, chaudhuri:index-selection, ai-meets-ai, selection-materialized-views, select-mviews}, which are complementary to our work. 
We leave the rigorous study of this problem to future work.

%% file: evaluation.tex
\section{Evaluation}
\label{sec:evaluation}
The goal of our experiments is two-fold. First, we demonstrate the tunability and space-efficiency of A+ indexes on three very different popular applications that GDBMSs support: (i) labelled subgraph queries; (ii) recommendations; and (iii) financial fraud detection. By either tuning the system's primary A+ index or adding secondary A+ indexes, we improve the performance of the system significantly, with minimal memory overheads. Second, we evaluate the performance and memory overhead tradeoffs of different A+ indexes on these workloads. Finally, as a baseline comparison, we
benchmark our performance against Neo4j~\cite{neo4j} and TigerGraph~\cite{tigergraph}, two commercial GDBMSs that have fixed adjacency list structures.
For completeness of our work, we also evaluate the maintenance performance of our indexes in the longer version of our paper~\cite{mhedhbi2020a}. 

\subsection{Experimental Setup}
\label{sec:experimental-setup}

We use a single machine 
with two Intel E5-2670 @2.6GHz CPUs and 512 GB of RAM. The machine has 16 physical cores and 32 logical 
ones.
Table~\ref{tab:datasets} shows the datasets used. We ran our experiments on all datasets and report numbers on a subset of datasets due to limited space. Our datasets include social, web, and Wikipedia knowledge graphs, which have a variety of graph topologies and sizes ranging from several million edges to over a hundred-million edges. A dataset G, denoted as G$_{i,j}$, has $i$ and $j$ randomly generated vertex and edge labels, respectively. We omit $i$ and $j$ when both are set to 1. We use query workloads drawn from real-world applications: (i) edge- and vertex-labelled subgraph queries; (ii) 
Twitter MagicRecs recommendation engine~\cite{gupta:magicrecs}; and (iii) fraud detection in financial networks.
For all index configurations (Configs), we report either the index reconfiguration (IR) or the index creation (IC) time of the newly added secondary indexes. 
All experiments use a single thread except the creation of edge-partitioned indexes, which uses 16 threads.

\begin{table}[t!]
	\centering
	\vspace{0.95em}
	\begin{tabular}{lrrc}\toprule
		Name             & \#Vertices & \#Edges   & Avg. degree\\
		\midrule
		Orkut (Ork)       &  3.0M & 117.1M & 39.03 \\
		LiveJournal (LJ)  &  4.8M &  68.5M & 14.27 \\
		Wiki-topcats (WT) &  1.8M &  28.5M & 15.83 \\
		BerkStan (Brk)    &  685K &   7.6M & 11.09 \\
		\bottomrule
	\end{tabular}
	\caption{Datasets used.}
	\label{tab:datasets}
\end{table}

\subsection{Primary A+ Index Reconfiguration}
\label{sec:reconfig}

\input{labeled-subgraph-experiment-tables.tex}

We first demonstrate the benefit and overhead tradeoff of tuning the primary A+ index in two different ways: (i) by only changing the sorting criterion; and (ii) by adding a new secondary partitioning. We used a popular subgraph query workload in graph processing that consists of labelled subgraph queries where both edges and vertices have labels.
We followed the data and subgraph query generation methodology from several prior work~\cite{mhedhbi:sqs, cfl}. We took the 14 queries from reference~\cite{mhedhbi:sqs} (omitted due to space reasons), which contain acyclic and cyclic queries with dense and sparse connectivity with up to 7 vertices and 21 edges. This query workload
had only fixed edge labels in these queries, for which GraphflowDB's default indexes are optimized. 
We modify this workload by also fixing vertex labels in queries. 
We picked the number of labels for each dataset to ensure that queries would take time in the order of seconds to several minutes.
Then we ran GraphflowDB on our workload on each of our datasets under three Configs:
\begin{enumerate}
\item \texttt{D}: system's default configuration, where edges are partitioned by edge labels and sorted by neighbour IDs.
\item\texttt{D$_s$}: keeps \texttt{D}'s secondary partitioning but  sorts edges first by neighbour vertex labels and then on neighbour IDs. 
\item \texttt{D$_{p}$}: keeps  \texttt{D}'s sorting criterion and edge label partitioning but adds a new secondary partitioning on neighbour vertex labels.
\end{enumerate}

\noindent Table~\ref{tab:labeled-sqs-runtime} shows our results. We omit Q14, which had very few or no output tuples on our datasets. 
First observe that \texttt{D$_{s}$} outperforms \texttt{D} on all of the 52 settings and by up to 10.38x and without any memory overheads as \texttt{D$_{s}$} simply changes the sorting criterion of the indexes. 
Next observe that by adding an additional partitioning level on \texttt{D}, the joins get even faster consistently across all queries, e.g., SQ$_{13}$ improves from 2.36x to 3.84x on Ork$_{8,2}$, as the system can directly access edges with a particular edge label and neighbour label using \texttt{D$_p$}. In contrast, under \texttt{D$_{s}$}, the system performs binary searches inside lists to access the same set of edges. Even though \texttt{D$_{p}$} is a reconfiguration, so does not index new edges, it still has minor memory overhead ranging from 1.05x to 1.15x because of the cost of storing the new partitioning layer. 
This demonstrates the effectiveness of tuning A+ indexes to optimize the system to be much more efficient on a different workload without any data remodelling, and with no (or minimal) memory overhead.

\subsection{Secondary Vertex-Partitioned A+ Indexes}
\label{subsec:vp-evaluation}

We next study the tradeoffs offered by secondary vertex-partitioned A+ indexes. We use two sets of workloads drawn from real-world applications that benefit from using both the system's primary A+ index as well as a secondary vertex-partitioned A+ index. Our two applications highlight two separate benefits users get from vertex-partitioned A+ indexes: (i) decreasing the amount of predicate evaluation; and (ii) allowing the system to generate new WCOJ plans that are not possible with only the primary A+ index.

\vspace{0.2em}
\subsubsection{Decreasing Predicate Evaluations}
\label{sec:predicate-avoidance}
In this experiment, we take a set of the queries drawn from the MagicRecs workload described in reference~\cite{gupta:magicrecs}. MagicRecs was a recommendation engine that was developed at Twitter that looks for the following patterns: for a given user $a_1$, it searches for users $a_2$...$a_k$ that $a_1$ has started following recently, and finds their common followers. These common followers are then recommended to $a_1$. We set $k$$=$2,3 and 4. Our queries, $MR_1$, $MR_2$, and $MR_3$ are shown in Figure~\ref{fig:magicrec-queries}.
These queries have a time predicate on the edges starting from $a_1$ which can benefit from indexes that sort on time. $MR_2$, and $MR_3$ are also cyclic can benefit from the default sorting order of the primary A+ index on neighbour IDs.
We evaluate our queries on all of our datasets on two Configs. The first Config consists of the system's primary A+ index denoted by \texttt{D} as before. The second Config, denoted by \texttt{D+VP$_t$} adds on top of D a new secondary vertex-partitioned index \texttt{VP$_{t}$} in the forward direction that: (i) has the same partitioning structure as primary forward A+ index so shares the same partitioning levels as the primary A+ index;  and (ii) sorts the inner-most lists on the \texttt{time} property of edges.
In our queries we set the value of $\alpha$ in the time predicate to have a 5\% selectivity.
For MR$_3$, on datasets LJ and Ork, we fix a$_1$ to 10000 and 7000 vertices, respectively, to run the query in a reasonable time.

Table~\ref{tab:magicrec-runtime} shows our results. First observe that despite indexing all of the edges again, our secondary index has only 1.08x memory overhead because: (i) the secondary index can share the partitioning levels
of the primary index; and (ii) the secondary index stores offset lists which has a low-memory footprint. In return, we see up to 10.6x performance benefits. We note that GraphflowDB uses exactly the same plans under both Configs that start reading $a_1$, extends to its neighbours and finally performs a multiway intersection (except for MR$_1$, which is followed by a simple extension). The only difference is that under \texttt{D+VP$_{t}$} the first set of extensions require fewer predicate evaluation because of accessing $a_1$'s adjacency list in \texttt{VP$_{t}$}, which is sorted on time. Overall this memory performance tradeoff demonstrates that with minimal overheads of an additional index, users obtain significant performance benefits on applications like MagicRecs that require performing complex joins for which the system's primary indexes are not tuned.

\input{magicrec-queries}

\input{magicrec-experiment-table}

\input{fraud-detection-queries}

\subsubsection{WCOJ Plans}
\label{sec:wco-query-plans}

We next evaluate the benefit and overhead tradeoff of secondary vertex-partitioned indexes when the secondary index  allows the system to generate new WCOJ plans that are not in the plan space 
with primary indexes only. We take a set of queries drawn from cyclic fraudulent money flows reported in prior literature~\cite{qiu:fraud-cycle}, as well as acyclic patterns that contain the money flow paths from our running examples. Figure~\ref{fig:fraud-app-queries} shows our queries MF$_1$, ..., MF$_5$. 
We focus on MF$_1$ to MF$_4$ here and use MF$_5$ in the next section. These four queries have equality conditions on the \texttt{city} property of the vertices, so can benefit from multiway joins computed by intersecting lists that are presorted on \texttt{city}. We evaluate these queries on two Configs. The first Config consists of the system's primary A+ index denoted by \texttt{D} as before. The second Config, denoted by \texttt{D+VP$_{c}$} adds on top of D a new secondary vertex-partitioned index \texttt{VP$_{c}$} in both forward and backward directions that: (i) has the same partitioning structure as primary A+ indexes; and (ii) sorts the inner-most lists on neighbour's \texttt{city} property.
For each dataset, we randomly added each vertex an account type property from [CQ, SV], a city from 4417 cities, and to each edge an amount in the range of [1, 1000] and a date within a 5 year range. 

Table~\ref{tab:fraud-sqs-runtime} shows our results (ignore the  MF$_5$ column and the \texttt{D+VP$_{c}$+EP$_{c}$} rows for now). Similar to our previous experiment, despite indexing all of the edges (this time twice), our secondary index has only 1.17x memory overhead (the increase from 1.08x is due to double indexing), whereas we see uniform and up to 24.7x improvements in run time. We note that in all of these queries, the benefits are solely coming from using new plans that use WCOJ processing. 
For example in query MF$_1$, the \texttt{D+VP$_{c}$} configuration allows the system to generate a plan that: (1) reads $a_1$; (2) uses \textsc{Multi-Extend} to intersect $a_1$'s forward and backward lists in \texttt{VP$_{c}$}, which matches $a_2$ and $a_4$; and (3) uses \textsc{E/I} that intersects $a_2$'s forward and $a_4$'s backward lists in the primary A+ index to match the $a_3$'s. Such plans are not possible in absence of the  \texttt{VP$_{c}$} index. Instead for  MF$_1$, under the default configuration D, the system extends $a_1$ to $a_2$, then to $a_3$ separately, runs a \textsc{Filter} operator to match the cities, and then uses \textsc{E/I} to match the $a_3$'s.

\subsection{Secondary Edge-Partitioned A+ Indexes}
\label{sec:secondary-edge-bound-eval}
Finally, we evaluate the tradeoffs of our secondary edge-partitioned A+ indexes on our financial fraud application.
We add a third Config to our experiment denoted by \texttt{D+VP$_{c}$+EP$_{c}$}. The configuration adds the edge-partitioned index from Example~\ref{ex:money-flow} in Section~\ref{subsec:secondary-eb-indexes}. We change the second-level partitioning to be on v.$_{adj}$.acc instead of edge labels and add the predicate e.$_{b}$.amt $<$ e.$_{nbr}$.amt + $\alpha$.
We pick the ``intermediate cut'' value $\alpha$ in our examples to have 5\% selectivity.

Table~\ref{tab:fraud-sqs-runtime} shows our results. First we observe that the addition of
\texttt{EP$_{c}$} only allows new plans to be generated for MF$_3$, MF$_4$ and MF$_5$, so we report numbers only for these queries. The improvements in run time range from 6.14x to up to 72.2x for a 2.22x memory overhead. Naturally the memory and performance tradeoff will change with the selectivity of $\alpha$. 
What is more important to note is that the speedups are primarily due to the system producing significantly more efficient plans in the presence of the \texttt{EP$_{c}$} index. 
For example, the system now generates a highly complex plan for MF$_3$, shown in Figure~\ref{fig:ex-wco-style-plan},
that uses a mix of vertex and edge-partitioned indexes and performs a 3-way intersection.

\input{example-plan}

\input{fraud-subgraph-experiment-tables}

\vspace{-0.1cm}
\subsection{Neo4j and TigerGraph Comparisons}
\label{app:baseline-comparisons}
We next compare GraphflowDB to Neo4j and TigerGraph. 
These experiments are provided for completeness only.
These are full-fledged commercial systems that support
transactions. However Neo4j is perhaps the most popular existing GDBMS
and TigerGraph, to the best of our knowledge, is the most performant one in terms of read performance. 
Our goal is to and show that the benefits of A+ indexes reported are on top of a system that is already competitive with existing GDBMSs. 

We report numbers for four of our labelled subgraph queries SQ$_1$, SQ$_2$, SQ$_3$, and SQ$_{13}$ on LJ$_{12,2}$ and WT$_{4,2}$ on Neo4j and TigerGraph, using their default Configs and using the \texttt{D} and \texttt{D$_p$} configurations from Section~\ref{sec:reconfig} for GraphflowDB. Table~\ref{tab:graphflows-comparison-systems} shows our results. We found GraphflowDB to be faster on all queries on the \texttt{D} configuration except for SQ$_{13}$ on WT$_{4,2}$. 
In addition, similar to our experiments from Table~\ref{tab:labeled-sqs-runtime}, the \texttt{D$_{p}$} configuration makes GraphflowDB even more performant. TigerGraph was the fastest system on SQ$_{13}$, which is a long 5-edge path. 
We cannot inspect the source code but we suspect for paths TigerGraph extends each distinct intermediate node only once and they only report pairs of reachable nodes. However, note that using the reconfigured index $D_p$, GraphflowDB outperforms TigerGraph on LJ$_{12,2}$ and closes the gap on WT$_{4,2}$. 

We note that system-to-system comparisons should not be interpreted as one system being superior to another.
What is more important is that neither of these systems has a mechanism for tuning through index reconfiguration or construction to close their performance gaps on join-heavy queries. 

\begin{table}[t!]
	\centering
	\captionsetup{justification=centering}
	\vspace{0.75em}
	\begin{tabular}{m{0.2cm} | m{0.5cm} m{0.5cm} m{0.5cm} m{0.5cm} | m{0.5cm} m{0.5cm} m{0.5cm} m{0.5cm}}\toprule
	& \multicolumn{4}{c}{LJ$_{12,2}$} & \multicolumn{4}{c}{WT$_{4,2}$} \\\midrule
	& SQ$_1$ & SQ$_2$ & SQ$_3$ & SQ$_{13}$ & SQ$_1$ & SQ$_2$ & SQ$_3$ & SQ$_{13}$ \\\midrule
	\texttt{D}\newline \texttt{D$_p$} \newline \texttt{TG}\newline	\texttt{N4}
	& \textbf{0.4}  \newline \textbf{0.4} \newline  2.5 \newline 29.3
	& 1.4  \newline \textbf{0.7} \newline 11.8 \newline 35.3 
	& 1.1  \newline \textbf{0.6} \newline 15.2 \newline 36.8
	& 31.3 \newline \textbf{6.0} \newline 30.5 \newline $TL$
	& 0.6   \newline \textbf{0.3} \newline  1.6\newline 1.65k
    & 4.6   \newline \textbf{2.1} \newline  7.1\newline 876
	& 5.5   \newline \textbf{3.1} \newline 10.2\newline 82.9
	& 767.5 \newline 235.7 \newline \textbf{29.5}\newline $TL$ \\\bottomrule
	\end{tabular}
	\caption{Runtime (secs) of GraphflowDB on Configs D and D$_p$ introduced in Section~\ref{sec:reconfig}, runtime of TigerGraph (TG), and runtime of Neo4j (N4). $TL$ indicates $>$30 mins.}
	\label{tab:graphflows-comparison-systems}
	\vspace{4pt}
\end{table}

\subsection{Index Maintenance Performance}
We next benchmark the maintenance speed of each type of A+ index on
a micro-benchmark. We report our numbers for two datasets LJ$_{2,4}$ and Brk$_{2,2}$. We load 50\% of the dataset from the MagicRec application and insert the remaining 50\% of the edges one at a time and evaluate the speed of 5 Configs, each requiring progressively more maintenance work:
(i) \texttt{D$_s$} has no partitioning and sorts by the the adjacent vertices IDs; 
(ii) \texttt{D$_p$} partitions each adjacency list on adjacent edges \texttt{label}; 
(iii) \texttt{D$_{ps}$} sorts each partition in \texttt{D$_p$} by the adjacent vertices IDs; (iv) \texttt{D$_{ps}$+VP$_t$} creates a secondary adjacency list index on the time property for \texttt{D$_{ps}$}; and finally (v) \texttt{D$_{ps}$+EP$_t$}: an edge bound adjacency list index with the same partitioning and sorting as \texttt{VP$_t$} for the query $v_s$$-$$[e_b]$$\leftarrow$$v_d$$-$$[e_{adj}]$$\rightarrow$$v_{adj}$ with predicate e$_{b}$.time < e$_{adj}$.time + $\alpha$ that has a 1\% selectivity.

We report our numbers for two datasets LJ$_{2,4}$ and Brk$_{2,2}$ using a single thread. We were able to maintain the following update rates per second (reported respectively for  LJ$_{2,4}$ and Brk$_{2,2}$): 1.203M and 2.108M for \texttt{D$_s$}, 1.024M and 1.892M for \texttt{D$_p$}, 1.081M and 1.832M for \texttt{D$_{ps}$}, 706K and 1.691M for \texttt{D$_{ps}$+VP$_t$}, and 41K and 110K for \texttt{D$_{ps}$+EP$_t$}. Our update rate gets slower with additional complexity but we are able to maintain insert rates of between 50-100k edges/s for our edge-partitioned index and between 706K-2.1M for our vertex-partitioned indexes. Note that our implementation is not write optimized and these speeds, though we believe is sufficient for modern applications, can be further improved.

%% file: labeled-subgraph-experiment-tables.tex
\begin{table*}[t!]
	\centering
	\captionsetup{justification=centering}
	\setlength{\tabcolsep}{4.5pt}
	\begin{tabular}{m{0.56cm} m{0.12cm} m{0.78cm} m{0.78cm} m{0.78cm} m{0.78cm} m{0.78cm} m{0.78cm} m{0.78cm} m{0.78cm} m{0.78cm} m{0.78cm} m{0.82cm} m{0.78cm} m{0.78cm} m{0.84cm} m{0.78cm}}
		\toprule
		& & SQ$_1$ & SQ$_2$ & SQ$_3$ & SQ$_4$ & SQ$_5$ & SQ$_6$ & SQ$_7$ & SQ$_8$ & SQ$_9$ & SQ$_{10}$ & SQ$_{11}$ & SQ$_{12}$ & SQ$_{13}$ & Mm & IR \\
	    \midrule
		Ork$_{8,2}$ &
		\texttt{D}\newline
		\texttt{D$_s$}\newline
		\newline
		\texttt{D$_p$}\newline
		& 1.68 \newline 0.91 \textbf{(1.85x)} \newline 0.68 \textbf{(2.48x)}
		& 5.47 \newline 3.12 \textbf{(1.75x)} \newline 2.61 \textbf{(2.10x)}
		& 3.66 \newline 2.04 \textbf{(1.79x)} \newline 1.35 \textbf{(2.71x)}
		& 1.30 \newline 1.19 \textbf{(1.09x)} \newline 0.97 \textbf{(1.34x)}
		& 1.58 \newline 1.05 \textbf{(1.50x)} \newline 0.77 \textbf{(2.05x)}
		& 1.45 \newline 1.22 \textbf{(1.19x)} \newline 0.60 \textbf{(2.44x)}
		& 1.73 \newline 1.33 \textbf{(1.30x)} \newline 1.30 \textbf{(1.33x)}
		& 2.49 \newline 1.51 \textbf{(1.65x)} \newline 1.46 \textbf{(1.71x)}
		& 0.95 \newline 0.77 \textbf{(1.23x)} \newline 0.60 \textbf{(1.25x)}
		& 17.74 \newline 4.89 \textbf{(3.63x)} \newline 3.89 \textbf{(4.56x)}
		& 7536.9 \newline 725.9 \textbf{(10.38x)} \newline 704.9 \textbf{(10.69x)}
		& 54.86 \newline 41.92 \textbf{(1.31x)} \newline 28.32 \textbf{(1.94x)}
		& 131.5 \newline 55.62 \textbf{(2.36x)} \newline 34.22 \textbf{(3.84x)}
		& 2778 \newline 2778 \textbf{(1.0x)} \newline 3106 \textbf{(1.12x)}
		& - \newline 38.90 \newline - \newline 27.71 \newline - \\
	    \midrule
	    LJ$_{2,4}$ &
	    \texttt{D}\newline
	    \texttt{D$_s$}\newline
	    \newline
	    \texttt{D$_p$}\newline
		& 1.47 \newline 1.45 \textbf{(1.01x)} \newline 1.04 \textbf{(1.41x)}
		& 7.87 \newline 6.22 \textbf{(1.27x)} \newline 5.18 \textbf{(1.52x)}
		& 6.46 \newline 5.42 \textbf{(1.19x)} \newline 4.64 \textbf{(1.39x)}
		& 1.69 \newline 1.49 \textbf{(1.13x)} \newline 1.09 \textbf{(1.55x)}
		& 1.59 \newline 1.51 \textbf{(1.05x)} \newline 0.98 \textbf{(1.62x)}
		& 1.60 \newline 1.52 \textbf{(1.05x)} \newline 1.08 \textbf{(1.48x)}
		& 1.91 \newline 1.40 \textbf{(1.36x)} \newline 1.07 \textbf{(1.79x)}
		& 3.35 \newline 2.39 \textbf{(1.40x)} \newline 1.85 \textbf{(1.81x)}
		& 4.07 \newline 2.82 \textbf{(1.44x)} \newline 2.26 \textbf{(1.80x)}
		& 41.54 \newline 28.07 \textbf{(1.48x)} \newline 25.86 \textbf{(1.61x)}
		& 807.8 \newline 241.2 \textbf{(3.35x)} \newline 235.63 \textbf{(3.43x)}
		& 397.1 \newline 268.6 \textbf{(1.48x)} \newline 235.85 \textbf{(1.68x)}
		& 468.8 \newline 259.2 \textbf{(1.81x)} \newline 161.82 \textbf{(2.90x)}
		& 1016 \newline 1016 \textbf{(1.0x)} \newline 1164 \textbf{(1.15x)}
		& - \newline 20.83 \newline - \newline 19.92 \newline - \\
		\midrule
		WT$_{4,2}$ &
		\texttt{D}\newline
		\texttt{D$_s$}\newline
		\newline
		\texttt{D$_p$}\newline
		& 0.61 \newline 0.37 \textbf{(1.65x)} \newline 0.32 \textbf{(1.91x)}
		& 4.59 \newline 2.43 \textbf{(1.89x)} \newline 2.09 \textbf{(2.20x)}
		& 5.48 \newline 3.50 \textbf{(1.56x)} \newline 3.05 \textbf{(1.80x)}
		& 0.84 \newline 0.69 \textbf{(1.22x)} \newline 0.55 \textbf{(1.53x)}
		& 1.17 \newline 0.71 \textbf{(1.65x)} \newline 0.59 \textbf{(1.99x)}
		& 0.90 \newline 0.65 \textbf{(1.38x)} \newline 0.54 \textbf{(1.66x)}
		& 0.73 \newline 0.61 \textbf{(1.20x)} \newline 0.61 \textbf{(1.21x)}
		& 11.25 \newline 3.93 \textbf{(2.87x)} \newline 2.86 \textbf{(3.94x)}
		& 2.85 \newline 1.36 \textbf{(2.09x)} \newline 1.09 \textbf{(2.62x)}
		& 1116.2 \newline 697.9 \textbf{(1.60x)} \newline 639.7 \textbf{(1.74x)}
		& 340.0 \newline 77.11 \textbf{(4.41x)} \newline 76.32 \textbf{(4.45x)}
		& 487.8 \newline 319.0 \textbf{(1.53x)} \newline 259.1 \textbf{(1.88x)}
		& 767.5 \newline 386.8 \textbf{(1.98x)} \newline 235.7 \textbf{(3.26x)}
		& 713 \newline 713 \textbf{(1.0x)} \newline 795 \textbf{(1.12x)}
		& - \newline 8.70 \newline - \newline 6.25 \newline - \\
		\bottomrule
	\end{tabular}
	\caption{Runtime (secs) and memory usage in MBs (Mm) evaluating subgraph queries using three different index configurations: \texttt{D}, \texttt{D$_s$}, and \texttt{D$_p$} introduced in Section~\ref{sec:reconfig}. We report index reconfiguration (IR) time (secs).}
	\vspace{10pt}
	\label{tab:labeled-sqs-runtime}
\end{table*}

%% file: magicrec-queries.tex
\begin{figure}[t!]
\vspace{-4pt}	
\centering
\captionsetup{justification=centering}
\begin{subfigure}[b]{0.29\linewidth}
\centering
\begin{tikzpicture}[scale=0.36, transform shape,->,>=stealth', shorten >=1pt, auto,node distance=2.2cm, thick, main node/.style={circle,draw,font=\sffamily\Huge\bfseries,minimum size=13mm}, text node/.style={font=\sffamily\Huge}]
	\node[main node] (1) {$a_1$};
	\node[main node] (2) [right of=1] {$a_2$};
	\node[main node] (3) [right of=2] {$a_3$};
	\path[every node/.style={font=\sffamily\Huge}]
	(1) edge node[xshift=0mm,yshift=2.6mm]{$e_1$} (2)
	(2) edge node[xshift=0mm,yshift=2.6mm]{$e_2$} (3) ;
\end{tikzpicture}
\vspace{-6pt}
\begin{lstlisting}[numbers=none, mathescape=true, showstringspaces=false]
 $P_\alpha$$($$e_1$$)$$,$$P_\alpha$$($$e_2$$)$
\end{lstlisting}
\vspace{-10pt}
\caption{MR$_1$.}
\label{fig:magicrec-Q1}
\end{subfigure}
\begin{subfigure}[b]{0.29\linewidth}
\centering
\begin{tikzpicture}[scale=0.36, transform shape,->,>=stealth', shorten >=1pt, auto,node distance=2cm, thick, main node/.style={circle,draw,font=\sffamily\Huge\bfseries,minimum size=13mm}, text node/.style={font=\sffamily\Huge}]
	\node[main node] (1) {$a_1$};
	\node[main node] (2) [above right of=1] {$a_2$};
	\node[main node] (3) [below right of=1] {$a_3$};
	\node[main node] (4) [below right of=2] {$a_4$};
	\path[every node/.style={font=\sffamily\Huge}]
	(1) edge node[xshift=0mm,yshift=0.8mm] {$e_1$} (2)
	(1) edge node[xshift=-9mm,yshift=-8.8mm] {$e_2$} (3)
	(2) edge (4)
	(3) edge (4) ;
\end{tikzpicture}
\vspace{-6pt}
\begin{lstlisting}[numbers=none, mathescape=true, showstringspaces=false]
 $P_\alpha$$($$e_1$$)$$,$$P_\alpha$$($$e_2$$)$
\end{lstlisting}
\vspace{-10pt}
\caption{MR$_2$.}
\label{fig:magicrec-Q2}
\end{subfigure}
\begin{subfigure}[b]{0.31\linewidth}
\centering
\hspace{0.2em}
\begin{tikzpicture}[scale=0.36, transform shape,->,>=stealth', shorten >=1pt, auto,node distance=2.2cm, thick, main node/.style={circle,draw,font=\sffamily\Huge\bfseries,minimum size=13mm}, text node/.style={font=\sffamily\Huge}]
	\node[main node] (1) {$a_1$};
	\node[main node] (3) [right of=1] {$a_3$};
	\node[main node] (2) [above of=3] {$a_2$};
	\node[main node] (4) [below of=3] {$a_4$};
	\node[main node] (5) [right of=3] {$a_5$};
	\path[every node/.style={font=\sffamily\Huge}]
	(1) edge node[xshift=0mm,yshift=0mm]{$e_1$} (2)
	    edge node[xshift=0mm,yshift=0mm]{$e_2$} (3)
	    edge node[xshift=-9mm,yshift=-7mm]{$e_3$} (4)
	(2) edge (5)
	(3) edge (5)
	(4) edge (5);
\end{tikzpicture}
\vspace{-6pt}
\begin{lstlisting}[numbers=none, mathescape=true, showstringspaces=false]
$P_\alpha$$($$e_1$$)$$,$$P_\alpha$$($$e_2$$)$$,$$P_\alpha$$($$e_3$$)$
\end{lstlisting}
\vspace{-10pt}
\caption{MR$_3$.}
\label{fig:magicrec-Q3}
\end{subfigure}
\caption{MagicRec (MR) queries. P$_\alpha$($e_i$) = $e_i$.time $<$ $\alpha$}
\label{fig:magicrec-queries}
\vspace{0.26em}
\end{figure}
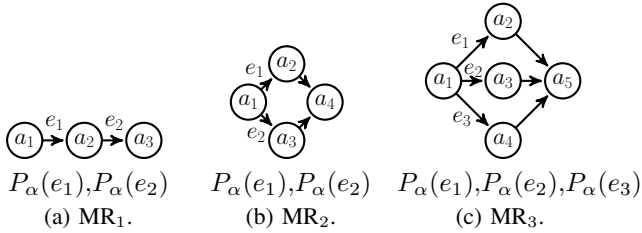

%% file: magicrec-experiment-table.tex
\begin{table}[t!]
	\centering
	\captionsetup{justification=centering}
	\setlength{\tabcolsep}{4.5pt}
	\begin{tabular}{m{0.24cm} m{0.55cm} m{1.2cm} m{1.2cm} m{1.3cm} m{1.2cm} m{0.6cm}}
		\toprule
		& & MR$_1$ & MR$_2$ & MR$_3$ & Mm & IC \\
		\midrule
		Ork &
		\texttt{D}\newline
		\texttt{D+VP$_t$}
		& 29.37 \newline 14.36\textbf{(2.0x)}
		& 255.4 \newline 166.3\textbf{(1.5x)}
		& 22.65 \newline 3.33\textbf{(6.8x)}
		& 2755 \newline 2982\textbf{(1.1x)}
		& - \newline 42.10 \\
		\midrule
		LJ &
		\texttt{D}\newline
		\texttt{D+VP$_t$}
		& 18.19 \newline 8.83\textbf{(2.1x)}
		& 38.17 \newline 27.26\textbf{(1.4x)}
		& 842.8 \newline 79.72\textbf{(10.6x)}
		& 1689 \newline 1820\textbf{(1.1x)}
		& - \newline 21.79 \\
		\midrule
		WT &
		\texttt{D}\newline
		\texttt{D+VP$_t$}
		& 6.87 \newline 2.69\textbf{(2.6x)}
		& 9.67 \newline 5.36\textbf{(1.8x)}
		& 136.5 \newline 22.74\textbf{(6.0x)}
		& 700 \newline 755\textbf{(1.1x)}
		& - \newline 9.14 \\
		\bottomrule
	\end{tabular}
	\caption{Runtime  (secs) and memory usage in MBs (Mm) evaluating MagicRec queries using Configs: \texttt{D} and \texttt{D+VP$_t$} introduced in Section~\ref{sec:predicate-avoidance}. We report index creation (IC) time (secs) for secondary indexes.}
	\label{tab:magicrec-runtime}
\end{table}

%% file: fraud-detection-queries.tex
\begin{figure}[t!]
	\centering
	\captionsetup{justification=centering}
\begin{subfigure}[b]{0.45\linewidth}
\centering
\begin{tikzpicture}[scale=0.36, transform shape,->,>=stealth', shorten >=1pt, auto,node distance=2.2cm, thick, main node/.style={circle,draw,font=\sffamily\Huge\bfseries,minimum size=13mm}, text node/.style={font=\sffamily\Huge}]
	\node[main node] (1) {$a_1$};
	\node[main node] (2) [right of=1] {$a_2$};
	\node[main node] (3) [above of=2] {$a_3$};
	\node[main node] (4) [left of=3]  {$a_4$};
	\path[every node/.style={font=\sffamily\Huge}]
	(1) edge node[xshift=0mm,yshift=0.9mm]{$e_1$} (2)
	(2) edge node[xshift=12mm,yshift=0mm]{$e_2$} (3)
	(3) edge node[xshift=0mm,yshift=9.6mm]{$e_3$} (4)
	(4) edge node[xshift=-12mm,yshift=0mm]{$e_4$} (1) ;
\end{tikzpicture}
\vspace{-5pt}
\begin{lstlisting}[numbers=none, mathescape=true, showstringspaces=false]
    $a_i$$.$$a$$c$$c$$=$$C$$Q$$,$
   $a_2$$.$$c$$i$$t$$y$$=$$a_4$$.$$c$$i$$t$$y$
\end{lstlisting}
\vspace{-10pt}
\caption{MF$_1$.}
\label{fig:fraud-Q1}
\end{subfigure}\hspace{3em}
\begin{subfigure}[b]{0.4\linewidth}
	\centering
	\begin{tikzpicture}[scale=0.36, transform shape,->,>=stealth', shorten >=1pt, auto,node distance=2.5cm, thick, main node/.style={circle,draw,font=\sffamily\Huge\bfseries,minimum size=13mm}, text node/.style={font=\sffamily\Huge}]
	\node[main node] (1) {$a_1$};
	\node[main node] (2) [right of=1] {$a_2$};
	\node[main node] (3) [right of=2] {$a_3$};
	\node[main node] (4) [right of=3] {$a_4$};
	\path[every node/.style={font=\sffamily\Huge}]
	(1) edge node[xshift=-1mm,yshift=1mm]{$e_1$} (2)
	(2) edge node[xshift=-1mm,yshift=1mm]{$e_2$} (3)
	(3) edge node[xshift=-1mm,yshift=1mm]{$e_3$} (4) ;
	\end{tikzpicture}
\begin{lstlisting}[numbers=none, mathescape=true, showstringspaces=false]
    $a_1$$.$$c$$i$$t$$y$$=$$a_2$$.$$c$$i$$t$$y$$,$
    $a_2$$.$$c$$i$$t$$y$$=$$a_3$$.$$c$$i$$t$$y$$,$
    $a_3$$.$$c$$i$$t$$y$$=$$a_4$$.$$c$$i$$t$$y$
\end{lstlisting}
\vspace{-10pt}
\caption{MF$_2$.}
\label{fig:fraud-Q2}
\end{subfigure}\vspace{-20pt}
\begin{subfigure}[b]{1\linewidth}
\centering
\begin{tikzpicture}[scale=0.36, transform shape,->,>=stealth', shorten >=1pt, auto,node distance=2.4cm, thick, main node/.style={circle,draw,font=\sffamily\Huge\bfseries,minimum size=13mm}, text node/.style={font=\sffamily\Huge}]
    \node[main node] (3) {$a_3$};
	\node[main node] (1) [left of=3]{$a_1$};
	\node[main node] (2) [above of=3] {$a_2$};
	\node[main node] (4) [right of=3] {$a_4$};
	\node[main node] (5) [below of=3] {$a_5$};
	\path[every node/.style={font=\sffamily\Huge}]
	(1) edge node{$e_1$} (2)
	    edge node[xshift=0mm,yshift=0.6mm]{$e_2$} (3)
		edge node[xshift=-8mm,yshift=-7mm]  {$e_4$} (5)
	(3) edge node[xshift=-1mm,yshift=0.6mm]{$e_3$} (4) ;
\end{tikzpicture}
\vspace{-6pt}
\begin{lstlisting}[numbers=none, mathescape=true, showstringspaces=false]
     $a_2$$.$$c$$i$$t$$y$$=$$a_4$$.$$c$$i$$t$$y$$,$  $a_4$$.$$c$$i$$t$$y$$=$$a_5$$.$$c$$i$$t$$y$$,$ $a_3$$.$$I$$D$$<$$1$$0$$0$$0$$0$$,$
         $a_i$$.$$a$$c$$c$$=$$C$$Q$$,$ $a_5$$.$$a$$c$$c$$=$$S$$V$$,$ $P_f$$($$e_2$$,$$e_3$$)$
\end{lstlisting}
\vspace{-10pt}
\caption{MF$_3$.}
\label{fig:fraud-Q3}
\end{subfigure}\vspace{6pt}
\begin{subfigure}[b]{1\linewidth}
	\centering
	\begin{tikzpicture}[scale=0.36, transform shape,->,>=stealth', shorten >=1pt, auto,node distance=2.5cm, thick, main node/.style={circle,draw,font=\sffamily\Huge\bfseries,minimum size=13mm}, text node/.style={font=\sffamily\Huge}]
	\node[main node] (1) {$a_1$};
	\node[main node] (2) [right of=1] {$a_2$};
	\node[main node] (3) [right of=2] {$a_3$};
	\node[main node] (4) [left of=1]  {$a_4$};
	\node[main node] (5) [left of=4]  {$a_5$};
	\path[every node/.style={font=\sffamily\Huge}]
	(1) edge node[xshift=0mm,yshift=1mm]{$e_1$} (2)
	    edge node[xshift=0mm,yshift=9mm]{$e_3$} (4)
	(2) edge node[xshift=0mm,yshift=1mm]{$e_2$} (3)
	(4) edge node[xshift=0mm,yshift=9mm]{$e_4$} (5) ;
	\end{tikzpicture}
\vspace{-5pt}
\begin{lstlisting}[numbers=none, mathescape=true, showstringspaces=false]
   $a_1$$.$$c$$i$$t$$y$$=$$\beta$$,$ $a_2$$.$$c$$i$$t$$y$$=$$a_4$$.$$c$$i$$t$$y$$,$ $a_2$$.$$a$$c$$c$$=$$C$$Q$$,$ $a_3$$.$$a$$c$$c$$=$$C$$Q$$,$
      $a_4$$.$$a$$c$$c$$=$$S$$V$$,$ $a_5$$.$$a$$c$$c$$=$$S$$V$$,$ $P_f$$($$e_1$$,$$e_2$$)$$,$ $P_f$$($$e_3$$,$$e_4$$)$
\end{lstlisting}
\vspace{-10pt}
\caption{MF$_4$.}
\label{fig:fraud-Q6}
\end{subfigure}\vspace{6pt}
\begin{subfigure}[b]{1\linewidth}
	\centering
	\begin{tikzpicture}[scale=0.36, transform shape,->,>=stealth', shorten >=1pt, auto,node distance=2.5cm, thick, main node/.style={circle,draw,font=\sffamily\Huge\bfseries,minimum size=13mm}, text node/.style={font=\sffamily\Huge}]
	\node[main node] (1) {$a_1$};
	\node[main node] (2) [right of=1] {$a_2$};
	\node[main node] (3) [right of=2] {$a_3$};
	\node[main node] (4) [right of=3]  {$a_4$};
	\node[main node] (5) [right of=4]  {$a_5$};
	\path[every node/.style={font=\sffamily\Huge}]
	(1) edge node[xshift=0mm,yshift=1mm]{$e_1$} (2)
	(2) edge node[xshift=0mm,yshift=1mm]{$e_2$} (3)
	(3) edge node[xshift=0mm,yshift=1mm]{$e_3$} (4)
	(4) edge node[xshift=0mm,yshift=1mm]{$e_4$} (5) ;
	\end{tikzpicture}
	\vspace{-5pt}
\begin{lstlisting}[numbers=none, mathescape=true, showstringspaces=false]
 $a_1$$.$$I$$D$$<$$5$$0$$0$$0$$0$$,$ $a_i$$.$$a$$c$$c$$=$$C$$Q$$,$ $P_f$$($$e_1$$,$$e_2$$)$$,$$P_f$$($$e_2$$,$$e_3$$)$$,$$P_f$$($$e_3$$,$$e_4$$)$
\end{lstlisting}
	\vspace{-10pt}
	\caption{MF$_5$.}
	\label{fig:fraud-Q5}
\end{subfigure}
    \vspace{-12pt}
    \caption{Fraud detection queries. P$_f$($e_i$,$e_j$) defined as\\ $e_i$$.$$d$$a$$t$$e$$<$$e_j$$.$$d$$a$$t$$e$$,$ $e_i$$.$$a$$m$$t$$>$$e_j$$.$$a$$m$$t$$,$ $e_i$$.$$a$$m$$t$$<$$e_j$$.$$a$$m$$t$$+$$\alpha$.}
    \label{fig:fraud-app-queries}
    \vspace{0.2em}
\end{figure}
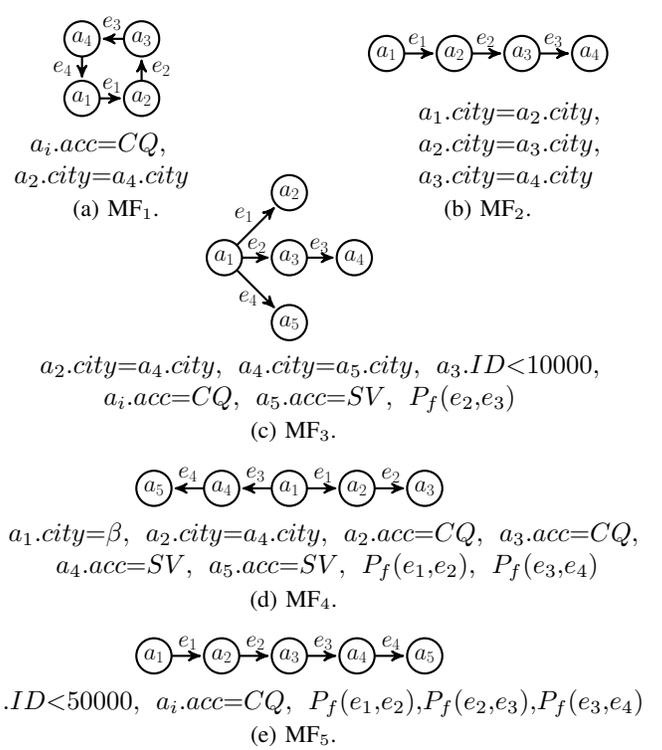

%% file: example-plan.tex
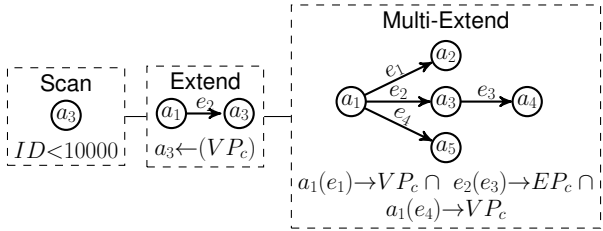
\begin{figure}[t!]
\captionsetup{justification=centering}
	\centering
	\vspace{-0.4em}
    \begin{tikzpicture}[grow=left]
    \node[draw,dashed,inner sep=\InnerSep] {
		\begin{tikzpicture}[scale=0.36, solid, transform shape,->,>=stealth',node distance=1.8cm, auto, thick, main node/.style={circle,draw,font=\sffamily\Huge\bfseries},text node/.style={font=\sffamily\Huge}]
			\node[main node] (1) {$a_1$};
			\node[main node] (2) [node distance =3.5cm,right of=1] {$a_3$};
			\node[main node] (3) [node distance =3cm,right of=2] {$a_4$};
			\node[main node] (4) [node distance =1.7cm, above of=2] {$a_2$};
			\node[main node] (5) [node distance =1.7cm,below of=2] {$a_5$};
			\node[text node] ()[node distance =1.32cm, above of=4]  {Multi-Extend};
            \node[text node](text1)[below of=5, node distance=1.3cm] {$a_1(e_1)$$\rightarrow$$VP_c$$ \ \cap $ \ $e_2(e_3)$$\rightarrow$$EP_c \ \cap $};
            \node[text node](text3)[below of=5, node distance=2.3cm] {$a_1(e_4)$$\rightarrow$$VP_c$};
			\path[every node/.style={midway, above,font=\sffamily\Huge}]
			(1) edge node[midway,above] {$e_{2}$} (2) edge node[sloped,midway,above] {$e_{1}$} (4) edge node[sloped, midway,above] {$e_{4}$} (5)
			(2) edge node {$e_{3}$} (3);
		\end{tikzpicture}
    }
    child [level distance=3.2cm] {node[draw,dashed,inner sep=\InnerSep] {
		\begin{tikzpicture}[scale=0.36,solid, transform shape,->,>=stealth', shorten >=1pt, auto,node distance=2.5cm, thick, main node/.style={circle,draw,font=\sffamily\Huge\bfseries},text node/.style={font=\sffamily\Huge}]
		\node[main node] (1) {$a_1$};
		\node[main node] (2) [right of=1] {$a_3$};
		\node[fit=(1) (2)] (cont1) {};
		\node[text node] ()[node distance =1.2cm,above of=cont1]  {Extend};
        \node[text node] () [below of=cont1,node distance=1.3cm] {$a_3$$\leftarrow$$(VP_c)$};
		\path[every node/.style={font=\sffamily\Huge,sloped}]
		(1) edge node[midway,above]{$e_{2}$} (2);
		\end{tikzpicture}
     }
     child [level distance=1.85cm]{node[draw,dashed,inner sep=\InnerSep] {
		\begin{tikzpicture}[scale=0.36,solid, transform shape,->,>=stealth', shorten >=1pt, auto,node distance=2cm, thick, main node/.style={circle,draw,font=\sffamily\Huge\bfseries},text node/.style={font=\sffamily\Huge}]
		\node[text node] ()[node distance =1.2cm,above of=1]  {Scan};
		\node[main node] (1) {$a_3$};
		\node[text node] () [below of=1,node distance=1.3cm] {$ID$$<$$1$$0$$0$$0$$0$};
		\end{tikzpicture}
     }
     }
     };
\end{tikzpicture}
    \caption{WCOJ Plan for MF$_3$ from Figure~\ref{fig:fraud-Q3} using two \texttt{VP$_c$} indexes and one \texttt{EP$_c$} index from Sections~\ref{sec:wco-query-plans} and~\ref{sec:secondary-edge-bound-eval}.}
\label{fig:ex-wco-style-plan}
	\vspace{3pt}
\end{figure}

%% file: fraud-subgraph-experiment-tables.tex
\begin{table*}[t!]
	\centering
	\captionsetup{justification=centering}
	\begin{tabular}{m{0.3cm} m{1.4cm} 
		m{1.4cm} m{1.4cm} m{1.4cm} m{1.4cm} m{1.4cm} m{1.4cm} m{1.2cm} m{0.6cm}}\toprule
		& & MF$_1$ & MF$_2$ & MF$_3$ & MF$_4$ & MF$_5$ & Mem(MB) & $|$E$_{indexed}$$|$ & IC \\\midrule
	    Ork &
	    \texttt{D}\newline
	    \texttt{D+VP$_c$}\newline
	    \texttt{D+VP$_c$+EP$_c$}
		& 73.35 \newline  8.99~\textbf{(8.16x)} \newline ---
		&  5.53 \newline  2.75~\textbf{(2.01x)}    \newline ---
		& 32.85 \newline  1.33~\textbf{(24.7x)} \newline 0.56~\textbf{(58.7x)}
		& 71.46 \newline 19.03~\textbf{(3.76x)} \newline 0.99~\textbf{(72.2x)}
		&  890.8 \newline --- \newline 60.59~\textbf{(14.7x)}
		&  2730 \newline 3183~\textbf{(1.17x)} \newline 6000~\textbf{(2.20x)}
		& 117.1M \newline 117.1M \newline 513.2M
		& - \newline 85.83 \newline 288.4 \\
		\midrule
		LJ &
		\texttt{D}\newline
		\texttt{D+VP$_c$}\newline
		\texttt{D+VP$_c$+EP$_c$}
		& 47.09 \newline 11.45~\textbf{(4.11x)} \newline ---
		&  4.24 \newline 2.86~\textbf{(1.48x)} \newline ---
		& 84.78 \newline 5.12~\textbf{(16.6x)} \newline 2.16~\textbf{(39.3x)}
		&  7.60 \newline 3.66~\textbf{(2.08x)} \newline 0.39~\textbf{(19.5x)}
		&  52.04 \newline  ---\newline 5.79~\textbf{(8.99x)}
		&  1649 \newline 1910~\textbf{(1.16x)} \newline 3585~\textbf{(2.17x)}
		& 68.5M \newline 68.5M \newline 276.2M
		& - \newline 46.43 \newline 279.8 \\
		\midrule
		WT &
		\texttt{D}\newline
		\texttt{D+VP$_c$}\newline
		\texttt{D+VP$_c$+EP$_c$}
		& 20.27 \newline 2.29~\textbf{(8.85x)} \newline ---
		&  1.47 \newline 1.12~\textbf{(1.31x)} \newline ---
		&  9.02 \newline 1.55~\textbf{(5.82x)} \newline 0.50~\textbf{(18.0x)}
		&  0.86 \newline 0.53~\textbf{(1.62x)} \newline 0.14~\textbf{(6.14x)}
		& 9.02 \newline --- \newline 0.79~\textbf{(11.4x)}
		&  685 \newline 796~\textbf{(1.16x)} \newline 1521~\textbf{(2.22x)}
		& 28.5M \newline 28.5M \newline 125.4M
		& - \newline 21.26 \newline 843.5 \\
		\bottomrule
	\end{tabular}
	\caption{Runtime (secs) of GraphflowDB plans and memory usage (Mem) in MB evaluating fraud detection queries using different Configs: \texttt{D}, \texttt{D+VP$_c$}, and \texttt{D+VP$_c$+EP$_c$} introduced in Section~\ref{sec:wco-query-plans}. The run time speedups and memory usage increase shown in parenthesis are in comparison to \texttt{D}. We report index creation time (IC) in secs for secondary indexes.}
	\label{tab:fraud-sqs-runtime}
	\vspace{10pt}
\end{table*}

%% file: related-work.tex
\vspace{2pt}
\section{Related Work}
\label{sec:related-work}

\noindent {\bf View-based Query Processing:} 
Answering queries using views has been well studied in the context of relational, XML, or RDF data management. We refer the reader to several surveys and references on the topic~\cite{halevy:survey, views-and-xml, materialized-views-rdf-data}.  
This extensive literature studies numerous topics, such as rewriting queries using a set of views~\cite{materialized-view-rdf-queries}, 
selecting a set of views for a workload e.g., web databases~\cite{view-selection-web-dbs}, or  the 
computational complexities of deciding whether a query can be answered 
with a given set of views~\cite{wenfei:use-views}. In this work, we observed that the lists that are stored in 
the adjacency list indexes can be seen as views and systems 
provide fast access to these lists/views through CSR-like data structures.
In contrast to prior work, we explored how to extend the views that can 
be accessed through adjacency list indexes in a space-efficient manner. 
Specifically, 
we identified a restricted but still much larger set of views than existing indexes, 
that can be stored by either merely tuning the partitioning schemes 
of a multi-level CSR data structure or lightweight offset lists. 

\noindent {\bf Kaskade~\cite{trindade:kaskade}} (KSK) is a graph query optimization framework 
that uses {\em materialized graph views} to speed up query evaluation. Specifically, KSK takes as input a query workload $\mathcal{Q}$ and an input graph $G$. Then, KSK enumerates possible {\em views} for $\mathcal{Q}$, which are other graphs $G'$ that contain a subset of the vertices in $G$ and other edges that can represent multi-hop connections in $G$. For example, if $G$ is a data provenance graph with job and file vertices, and there are ``consumes'' and ``produces'' relationships between jobs and files, an example graph view $G'$ could store the job vertices and their 2-hop dependencies through files. 
KSK materializes its selected views in Neo4j, and then translates queries over $G$ to appropriate graphs (views) that are stored in Neo4j, which is used to answer queries. Therefore, the framework is limited by Neo4j's adjacency lists. 

There are several differences between the views provided by KSK and A+ indexes.
First, KSK's views are based on ``constraints'' that are mined from $G$'s schema
based only on vertex/edge labels and not properties. 
For example, KSK can mine ``job vertices connect to jobs in 2-hops but not to file vertices'' constraints but not ``accounts connect to accounts in 2-hops with later dates and lower amounts'', which can be a predicate in an A+ index. 
Second, because KSK stores its views in Neo4j, 
KSK views are only vertex ID and edge label partitioned, unlike our views which are stored in 
a CSR data structure that support tunable partitioning, including by edge IDs, as well as sorting. Similarly, because KSK uses Neo4j's query processor, its plans do not use WCOJs.

\noindent {\bf Adjacency List Indexes in Graph Analytics Systems:}
There are numerous graph analytics systems~\cite{buluc:combinatorial-blas, 
malewicz:pregel, shun:ligra} 
designed to do batch analytics, such as decomposing a graph into connected components. These systems use native graph storage formats, such as adjacency lists or sparse matrices. Work in this space generally focuses on optimizing the physical layout of the edges in memory. 
For systems storing the edges in adjacency list structures, a common technique is to store them in CSR format~\cite{bonifati:graphs-book}. To implement A+ indexes we used a variant of CSR that can have multiple partitioning levels. 
Reference~\cite{shun:ligra} studies CSR-like partitioning techniques for large lists
and reference~\cite{zhang:cagra} proposes segmenting a graph stored in a CSR-like format 
for better cache locality. This line of work is complementary to ours. 

\noindent {\bf Indexes in RDF Systems:}
RDF systems support the RDF data model, in which data is represented as a set of (subject, predicate, object) triples. Prior work has introduced different architectures, such as storing and then indexing 
one large triple table~\cite{neumann:rdf,weiss:hexastore} or adopting a native-graph storage~\cite{zou:gstore}. 
These systems have different designs to further index these tables or their adjacency lists. For example,
RDF-3X~\cite{neumann:rdf} indexes an RDF dataset in multiple B+ tree indexes.
As another example, the gStore system encodes several vertices in fixed length bit strings that captures information about the neighborhoods of vertices. 
Similar to the GDBMSs we reviewed, these work also define fixed indexes for RDF triples.  A+ indexes instead gives users a tunable mechanism to tailor a GDBMS to the requirements of their workloads.

\noindent {\bf Indexes for XML Data:} 
There is prior work focusing on indexes for XML and the precursor tree or rooted graph data models.
Many of this work provides complete indexes, such as DataGuides~\cite{dataguides} or IndexFabric~\cite{index-fabric}, or approximate indexes~\cite{suciu, akindex} that index the paths from the roots of a graph to individual nodes in the data. These indexes are effectively summaries of the graph that are used during query evaluation to prune the search of path expressions in the data. These indexes are not directly suitable for contemporary GDBMS which store non-rooted property graphs, where the paths that users search in queries can start from arbitrary nodes.

\noindent {\bf Other complex subgraph indexes:} 
Many prior algorithmic work on evaluating subgraph queries~\cite{europar:path-index, cheng:fast-graph, lai:seed} have also proposed auxiliary indexes that index subgraphs more complex than edges, such as paths, stars,
or cliques. This line of work effectively
demonstrates that indexing such subgraphs can speed up subgraph
query evaluation.
Unlike our work, these subgraphs can be more complex but their storage is not optimized for space efficiency.

%% file: future-work.tex
\section{Conclusions}
\label{sec:future-work}
Ted Codd, the inventor of the relational model, criticized the GDBMSs of the time as being restrictive because they only performed a set of ``predefined joins''~\cite{ted}, which causes physical data dependence and contrasts with relational systems that can join arbitrary tables. This is indeed still true to a good extent for contemporary GDBMSs, which are designed to join vertices with only their neighbourhoods, which are predefined to the system as edges. However, this is specifically the major appeal of GDBMSs, which are highly optimized to perform these joins efficiently by using adjacency list indexes. 
Our work was motivated by the shortcoming that existing GDBMSs have fixed adjacency list indexes that limit the workloads that can benefit from their fast join capabilities. As a solution, we described the design and implementation of a new indexing subsystem with restricted materialized view support that can be stored using a space-efficient technique. We demonstrated the flexibility of A+ indexes, and evaluated the performance and memory tradeoffs they offer on a variety of applications drawn from popular real-world applications.\vspace{0.8em}